\documentclass[12pt]{article}

\usepackage{amsmath}
\usepackage{graphicx}
\usepackage{enumerate}
\usepackage[natbib=true, style=apa]{biblatex}
\usepackage{url} 

\newcommand{\blind}{0}

\newif\ifappendix
\appendixtrue
\newif\ifjasa
\jasafalse

\usepackage{graphicx}
\usepackage{amssymb}
\usepackage{mathtools}
\usepackage{epstopdf}
\usepackage[multiple]{footmisc}
\DeclarePairedDelimiter\norm{\lVert}{\rVert}
\DeclarePairedDelimiter\abs{\lvert}{\rvert}
\DeclarePairedDelimiter\set{\{}{\}}
\DeclarePairedDelimiter\p{(}{)}
\DeclarePairedDelimiter\sqb{[}{]}
\DeclarePairedDelimiter\inner{\langle}{\rangle}
\DeclareMathOperator{\vectorspan}{span}
\usepackage{bbm}
\usepackage{amssymb}
\usepackage{amsmath}
\usepackage{amsfonts}

\usepackage{lscape}
\usepackage{rotating}
\usepackage{setspace}
\usepackage{threeparttable}
\usepackage{booktabs}
\usepackage{titlesec} 
\usepackage{lmodern}
\fontfamily{lmtt}\selectfont
\usepackage[T1]{fontenc}

\DeclareMathOperator{\argmin}{argmin}

\RequirePackage[colorlinks,allcolors=blue]{hyperref}
\usepackage{titling}
\newcommand{\subtitle}[1]{%
\posttitle{%
\par\end{center}
\begin{center}\large#1\end{center}
\vskip0.5em}%
}
\usepackage{amsthm}
\theoremstyle{definition}
\newtheorem{theorem}{Theorem}[section]

\newtheorem{proposition}[theorem]{Proposition}

\newtheorem{assumption}{Assumption}

\usepackage{color}

\newcommand\indep{\protect\mathpalette{\protect\independenT}{\perp}}
\def\independenT#1#2{\mathrel{\rlap{$#1#2$}\mkern2mu{#1#2}}}
\newcommand{\R}{\ensuremath{\mathbb{R}}}

\newcommand{\calB}{\ensuremath{\mathcal{B}}}

\newcommand{\E}{\ensuremath{\mathbb{E}}}

\newcommand{\calS}{\ensuremath{\mathcal{S}}}

\newcommand{\calM}{\ensuremath{\mathcal{M}}}

\newcommand{\calH}{\ensuremath{\mathcal{H}}}

\newcommand{\Var}{\text{Var}}

\newcommand{\ipw}{\text{ipw}}

\newcommand{\estimand}{\psi}
\newcommand{\dispersion}{\chi}
\newcommand{\imbalancescale}{\zeta}

\newcommand{\dgamma}{\delta\gamma}
\newcommand{\dm}{\delta m}

\DeclareMathOperator{\imbalance}{imbalance}
\DeclareMathOperator{\var}{Var}

\newcommand{\Pn}{\ensuremath{P_n}}
\newcommand{\hgamma}{\ensuremath{\hat\gamma}}
\newcommand{\hgammaipw}{\ensuremath{\hat\gamma}}
\newcommand{\gammaipw}{\ensuremath{\gamma^{\ipw}}}
\newcommand{\model}{\mathcal{M}}

\newcommand{\aug}{\text{aug}}

\def\b1{\boldsymbol{1}}

\usepackage{fullpage}

\begin{document}

\def\spacingset#1{\renewcommand{\baselinestretch}%
{#1}\small\normalsize} \spacingset{1}

\if0\blind
{
\title{\bf The Balancing Act in Causal Inference\thanks{For comments and suggestions, we thank Peng Ding, Vitor Hadad, and Stefan Wager.} \bigskip}
\author{Eli Ben-Michael\thanks{Harvard University}
\hspace{-3.5mm} \and \hspace{-3.5mm}
Avi Feller\thanks{UC Berkeley}
\hspace{-3.5mm} \and \hspace{-2.5mm}
David A. Hirshberg\thanks{Emory University} 
\hspace{-3.5mm} \and \hspace{-3.5mm} 
Jos\'e R. Zubizarreta\thanks{Harvard University}
}
  \maketitle
} \fi

\if1\blind
{
  \bigskip
  \bigskip
  \bigskip
  \begin{center}
    {\LARGE\bf The Balancing Act in Causal Inference}
\end{center}
  \medskip
} \fi

\bigskip

\begin{abstract}
\singlespacing
The idea of covariate balance is at the core of causal inference. 
Inverse propensity weights play a central role because they are the unique set of weights that balance the covariate distributions of different treatment groups. 
We discuss two broad approaches to estimating these weights: 
the more traditional one, 
which fits a propensity score model and then uses the reciprocal of the estimated propensity score to construct weights,
and the balancing approach, which estimates the inverse propensity weights  essentially by the method of moments, finding weights that achieve balance in the sample.
We review ideas from the causal inference, sample surveys, and semiparametric estimation literatures, with particular attention to the role of balance as a sufficient condition for robust inference.  We focus on the inverse propensity weighting and augmented inverse propensity weighting estimators for the average treatment effect given strong ignorability and consider generalizations for a broader class of problems including policy evaluation and the estimation of individualized treatment effects.

\end{abstract}

\noindent%
\vfill

\newpage
\ifjasa
\spacingset{1.9} 
\else 
\onehalfspacing
\fi

\section{Introduction}
\label{sec:intro}

The idea of covariate balance is at the core of causal inference.
Intuitively, when estimating the effect of a treatment, we want to compare ``like with like'': if two groups are similar in every respect except in assignment to treatment, then we can attribute differences in outcomes to the treatment itself, as opposed to other confounding factors \citep{cochran1965planning}.

This intuition is widespread. However, any two groups will differ in some respect, so when we talk about a like-with-like comparison, 
we rely on a consensus about which 
differences matter and which can be ignored.  
For example, we might focus on imbalances in the means of covariates, or in their marginal distributions, or in summaries of their joint distribution. In this paper,
we explore the role covariate balance plays in statistical inference about treatment effects; its relationship to common notions of balance; and connections to the causal inference, sample surveys, and semiparametrics literatures. 

\emph{Inverse propensity weights} play a central role because they are the unique set of weights that balance the covariate distributions of different treatment groups.
However, outside randomized experiments, these weights are unknown and must be estimated. Different approaches to estimating it tend to arise from emphasis on different sufficient conditions for a successful estimate.

The most common approach to estimating inverse propensity weights is to estimate the conditional probability of treatment, or propensity score, and then take its reciprocal. 
With some estimation methods, this approach can be unstable, especially when this probability is close to zero, and can lead to poor balance in the sample at hand when the propensity score is not estimated accurately. 
Researchers often address these limitations with post-hoc fixes, e.g., truncating the extreme weights that arise from propensity score estimates that are too near zero or checking balance and re-estimating the propensity score if it is poor. 

An alternative approach is to find weights that optimize balance in the data at hand. 
Because balance is a defining property of the 
inverse propensity weights, this is essentially a method of moments estimate of the inverse propensity weights.
Researchers typically estimate such weights by solving a constrained optimization problem, e.g., by finding weights that restrict the maximum imbalance in covariate means between groups. 
Balance checks are therefore built in to this approach: we choose weights to pass the checks, rather than performing the checks post-hoc.
Examples of proposals in line with this approach include
\citet{athey2018approximate}, 
\citet{breidt2017model},
\citet{chan2016globally},
\citet{deville1992calibration},
\citet{graham2012inverse},
\citet{hainmueller2012balancing},
\citet{hirshberg2018augmented},
\citet{imai2014covariate}, 
\citet{kallus2020generalized},
\citet{li2018balancing},
\citet{tan2020regularized},
\citet{wong2018kernel},  
\citet{zhao2019covariate}, and
\citet{zubizarreta2015stable}.

We will organize our discussion around a specific form of balance that is necessary and sufficient for straightforward inference based on the widely-used inverse probability weighting (IPW) and augmented inverse probability weighting (AIPW) estimators. We need balance in the mean outcome conditional on treatment and covariates for IPW, and in the errors we make when estimating it for AIPW. 
This happens when we estimate the inverse propensity weights and this conditional mean function sufficiently well, an idea that gives rise to the concept of double-robustness, and also when we balance every function in a set that contains 
the conditional mean, an idea that gives rise to balance checks and weights optimized to achieve balance in this sense.  
In particular, many widely-used balance checks for weights are based on the \emph{maximal imbalance} for conditional means in a set encoding structural assumptions about the conditional mean. 
This gives us a practical framework to link intuition and assumptions about the data to appropriate balance criteria to check or optimize.


In Sections~\ref{sec:framework} and \ref{sec:ipw} respectively, we describe how balance controls the bias of the IPW and AIPW estimators and discuss the  propensity-modeling and balancing approaches to weighting. In Section~\ref{sec:asymptotic}, we characterize the asymptotic behaviour of the IPW and AIPW estimators in terms of balance. In Section~\ref{sec:balance_models}, we focus on the balancing approach, discussing the way we encode our beliefs about the data in our choice of what to balance and interpreting some balance criteria that are commonly used in practice. In Section \ref{sec:trade-offs}, we discuss the  inferential implications of the trade-off between the expressiveness of the set of functions we choose to balance and the extent to which we can achieve balance for all of these functions. In Section \ref{sec:misfit_toys}, we comment on additional practically important choices, e.g., whether to use weights that extrapolate and how to tune a bias-variance trade-off parameter common to many instances of balancing weights.
In Section~\ref{sec:other-estimands}, we generalize our discussion from
the specific problem of estimating average treatment effects to a broader class of
problems including policy evaluation and the estimation of individualized treatment effects. 
In Section~\ref{sec:dual}, we conclude with a discussion of the equivalence between balance-focused optimization problems and loss functions for estimating the propensity score, including equivalences or near-equivalences between specific approaches described as focused on balance and those described as focused on propensity estimation.

\section{Framework}
\label{sec:framework}

\vspace{-.25cm}
\subsection{Setup}

We posit the existence of independent and identically distributed tuples $(X_i, W_i, Y_i(0), Y_i(1))$ randomly drawn from a population, where $X_i \in \R^d$ are observed covariates, $W_i \in \{0,1\}$ is a binary treatment indicator, and $Y_i(0), Y_i(1) \in \R$ are the potential outcomes under control and treatment, respectively \citep{neyman1923,rubin1974estimating}.
Under the Stable Unit Treatment Value Assumption (SUTVA; \citeauthor{rubin1980}, 1980), we can write the observed outcome $Y_i$ as $Y_i = (1-W_i)Y_i(0) + W_i Y_i(1)$.
Since the tuples are randomly drawn from a population and we are interested in population-level estimands, we generally drop the subscript $i$.

We are primarily interested in the {\it Average Treatment Effect} (ATE), defined as
\begin{equation}
  \label{eq:pate}
  \tau := \E[Y(1) - Y(0)].
\end{equation}
To ease exposition, we separate it into two components, $\mu_1 := \E[Y(1)]$ and $\mu_0 := \E[Y(0)]$, and focus on estimating the treatment-specific mean $\mu_1$; we can estimate $\mu_0$ analogously.

We base our analysis on the assumption of strong ignorability \citep{Rosenbaum1983}.
\begin{assumption}[Ignorability]
  \label{a:ignore}
  $W \indep (Y(0), Y(1))\mid X$
\end{assumption}

\begin{assumption}[Overlap]
  \label{a:overlap}
  $e(x) > 0$ where $e(x) := \E[W \mid X = x]$ is the \emph{propensity score}
\end{assumption}
\noindent For some results, it is useful to consider a stronger overlap assumption, which rules out the possibility that rare treatment assignments have an undue influence on the estimand.\footnote{
This assumption is equivalent to a continuity property: if two conditional mean outcome functions $m$ and $\tilde m$ are close in the sense that $\E (m_W(X)-\tilde m_W(X))^2$
is small, then the corresponding treatment-specific means $\mu_1=\E[m_1(X)]$ and $\tilde\mu_1 = \E[\tilde m_1(X)]$ must be close as well. Without Assumption \ref{a:overlap} or strong assumptions on the (semi)parametric form of $m_w(x)$, there does not exist a regular estimator of $\mu_1$ \citep{van1991differentiable}. 
The stronger assumption of \emph{strict overlap}, 
that $e(x) \ge \eta$ for some $\eta > 0$, is common.}
{\addtocounter{assumption}{-1}%
   \renewcommand{\theassumption}{\arabic{assumption}$'$}
\begin{assumption}[Sufficient Overlap]
\label{a:regular-overlap}
$\E[1 / e(X)] < \infty.$
\addtocounter{theorem}{-1}
\end{assumption}}

\noindent Under Assumptions \ref{a:ignore} and \ref{a:overlap}, the treatment-specific mean is an average of the conditional mean outcome,
\begin{equation}
 \label{eq_identification}
\mu_1 = \E[m_1(X)] \quad \text{ where } \quad m_w(x) := \E[Y \mid X = x, W=w].
 \end{equation}
The main challenge when estimating $\mu_1$ is that, outside of randomized experiments, 
the mix of units that receive one treatment tends to differ from the mix that receive another.
Thus, the average outcome among those who receive the treatment reflects selection into treatment, as well as the treatment itself. 
Under Assumption \ref{a:ignore}, the observed covariates explain the salient differences between these groups, 
so eliminating selection effects reduces to a covariate adjustment problem.

\vspace{-.25cm}
\subsection{Inverse propensity weighting}
\label{sec:true_ipw}

A celebrated result in causal inference is that the inverse propensity weighted average of the outcomes is an unbiased estimator of $\mu_1$ under Assumptions \ref{a:ignore} and \ref{a:overlap} \citep{rosenbaum1987model}:
\begin{equation}
  \label{eq_mipw-identity}
  \mu_1 = \E\left[m_1(X_i)\right] =  \E\left[\frac{W_i}{e(X_i)} m_{W_i}(X_i)\right] = \E\left[\frac{W_i}{e(X_i)} Y_i \right].
\end{equation}

\noindent In fact, this result holds more generally: the inverse propensity weights are the unique weighting function that, for \emph{every} function $f(x)$, 
equates weighted averages of $f$ over the treated units to unweighted averages over the study population.
That is, $\gammaipw(x)=1/e(x)$ is the only function  satisfying
\begin{equation}
    \label{eq_mipw-balance}
    \E\left[W_i \gammaipw(X_i) f(X_i)\right] = \E\left[f(X_i)\right]  \ \text{ for all bounded }\ f(x). \end{equation}
Because the solution is unique, Equation~\ref{eq_mipw-balance} also \emph{defines} the inverse propensity weights $\gammaipw$.
We call Equation~\ref{eq_mipw-balance} the \emph{population balance property}: inverse propensity weights adjust for 
 differences (in the averages of all functions $f$ of the covariates) between the subpopulation receiving treatment and the overall population.
Setting $f=m_1$ recovers the unbiasedness result above \eqref{eq_mipw-identity}. The law of large numbers implies a corresponding \emph{sample balance property}:
\begin{equation}
    \label{eq_mipw-sample-balance}
    \frac{1}{n}\sum_{i=1}^n W_i \gammaipw(X_i) f(X_i) \approx \frac{1}{n}\sum_{i=1}^n f(X_i) \   \text{ for any bounded }\ f(x).
\end{equation}
This holds, for example, for the coordinate functions $f(x)=x_{j}$. When using an estimate of the inverse propensity weights $\gammaipw$, it is common to check that sample balance indeed holds, i.e., that each covariate has a similar average in the weighted treatment group and in the whole sample.

\vspace{-.25cm}
\subsection{Imbalance and estimation error}

As a starting point, we write out the estimation error for a general weighting estimator,
\begin{equation}
\label{eq_general_ipw}
\hat{\mu}_1 = \frac{1}{n}\sum_{i=1}^n W_i \hat{\gamma}(X_i) Y_i.
\end{equation}
Throughout, we will consider only  design-based weights $\hat{\gamma}(X_i)$, i.e., weights that depend on the study's \emph{design} --- the treatment statuses $W_i$ and covariates $X_i$ of all units --- but not on the outcomes $Y_i$.\footnotemark\
The error of this estimator is, in terms of $\varepsilon_i :=  Y_i - m_{W_i}(X_i)$,
\begin{equation}
    \label{eq:weights_error}
    \hat{\mu}_1 - \mu_1 = \underbrace{\frac{1}{n}\sum_{i=1}^nW_i\hat{\gamma}(X_i)  m_1(X_i) - \frac{1}{n}\sum_{i=1}^n m_1(X_i)}_{\text{imbalance in $m_1$}} + \underbrace{\frac{1}{n}\sum_{i=1}^n W_i\hat{\gamma}(X_i) \varepsilon_i}_{\text{noise}} + \underbrace{\frac{1}{n}\sum_{i=1}^{n} m_1(X_i) - \mu_1}_{\text{sampling variation}},
\end{equation}
 The first term in Equation~\ref{eq:weights_error} is the imbalance in the conditional mean $m_1(X_i)$. The second and third terms are a weighted average of noise and irreducible sampling variation respectively. 
Since these two terms both have mean zero conditional on the design, the design-conditional bias of the estimator \emph{is} the imbalance in $m_1$. \\

\footnotetext{
Furthermore, as implied by our notation, we will consider only weights that are functions of the covariate $X_i$.
This is almost universal among design-based weights, there being no design-based reason to distinguish units with identical covariate and treatment status by assigning them different weights, but is not always indicated notationally the way we do here: some write $\hgamma_i$ instead of $\hgamma(X_i)$ for the weight of the $i$th unit.}
 
What makes estimation challenging is that the conditional mean function $m_1$ is unknown. One natural approach is to think of $m_1$ as belonging to some set of functions $\model$, e.g., the set of linear functions, which serves as a \emph{model}.
When $m_1$ belongs to this set, the absolute design-conditional bias  
is bounded by the largest value it could take on for any $m_1 \in \model$, which we call the \emph{maximal imbalance}, $\imbalance_\model(\hgamma)$:
\begin{equation}
    \label{eq:bias}
    \abs*{\text{design-conditional bias}} \leq \imbalance_{\calM}(\hgamma) := \max_{m_1 \in \calM} \abs*{\frac{1}{n}\sum_{i=1}^n m_1(X_i) - \frac{1}{n}\sum_{i=1}^n W_i \hgamma(X_i) m_1(X_i) }.
\end{equation}
Thus, controlling maximal imbalance --- if we do a good job of choosing \emph{what} to balance --- is an effective way to control bias. 


\vspace{-.25cm}
\subsection{Augmented Inverse Propensity Weighting}
\label{sec:aipw}

A prominent generalization of the IPW estimator augments it with an adjustment involving an estimate of the conditional mean function $m_1$, 
e.g., fit via regression on the treated sample (Robins et al., \citeyear{robins1994estimation}). 
This augmented IPW (AIPW) estimator uses the regression estimator $\hat m_1$ to impute and adjust for any imbalance in $m_1$ that is left over after weighting:
\begin{equation}
  \label{eq:aipw2}
\hat{\mu}^{\text{aug}}_1 := \underbrace{\frac{1}{n}\sum_{i=1}^n W_i\hat{\gamma}(X_i) Y_i}_{\text{weighting estimator}} + \underbrace{\frac{1}{n}\sum_{i=1}^n \left\{\hat m_1(X_i) - W_i\hat{\gamma}(X_i)\hat m_1(X_i)\right\}}_{\text{bias correction via imputation}}.
\end{equation}
Though it is traditional to use a modeling approach to estimate the weights used in the AIPW estimator, 
many recent proposals combine this bias correction with control of maximal imbalance. Examples include \emph{approximate residual balancing} \citep{athey2018approximate}, \emph{augmented minimax linear estimation} \citep{hirshberg2018augmented}, and \emph{automatic double machine learning} \citep*{chernozhukov2018automatic}.

The augmented estimator has an error decomposition analogous to that of a pure weighting estimator. In it, imbalance in the \emph{regression error function}
$\delta m_1 = \hat m_1 - m_1$ replaces imbalance in $m_1$ itself.
\begin{equation}
    \label{eq:weights_error_aug}
    \hat{\mu}^{\aug}_1 - \mu_1 = \underbrace{\frac{1}{n}\sum_{i=1}^n \delta m_1(X_i) - \frac{1}{n}\sum_{i=1}^nW_i\hat{\gamma}(X_i) \delta m_1(X_i)}_{\text{imbalance in $\dm$}} + \underbrace{\frac{1}{n}\sum_{i=1}^n W_i\hat{\gamma}(X_i) \varepsilon_i}_{\text{noise}} + \underbrace{\frac{1}{n}\sum_{i=1}^{n} m_1(X_i) - \mu_1}_{\text{sampling variation}}.
\end{equation}
Intuitively, imbalance with augmentation should be smaller than without because the regression error function $\delta m$ should smaller than the conditional mean function $m_1$ itself.
This generalizes our previous error decomposition for $\hat\mu_1$: the IPW estimator is AIPW with $\hat m_1=0$, in which case $\delta m=-m_1$ and the error decomposition reduces to Equation~\ref{eq:weights_error}. 
Thus, maximal imbalance can play a similar role in our thinking; 
however, in this case we will want our model $\model$ to contain the regression error $\dm$, not $m_1$ itself.
For brevity, our discussion will be in terms of the more general augmented estimator and the regression error $\dm$, 
noting when we require that the regression estimator $\hat m_1$ is or is not zero. 

\section{Estimating inverse propensity score weights}
\label{sec:ipw}

In observational studies, the inverse propensity weights $\gammaipw$ are unknown and must be estimated. The two equivalent definitions of
$\gammaipw$, via the explicit form $\gammaipw(x)=1/e(x)$ and via the population balance property \eqref{eq_mipw-balance},
suggest two broad types of inverse propensity estimators.
Traditional modeling approaches estimate $e(x)$ by regressing treatment status on covariates.
Balancing estimators instead directly estimate weights $\gamma(x)$ that satisfy the sample balance property \eqref{eq_mipw-sample-balance}. 
We turn to these now.

\vspace{-.25cm}
\subsection{Modeling approach to weighting}
\label{sec:mipw}
By far, the most common approach to weighting is to take an estimate of the propensity score, $\hat{e}(x)$, and use its reciprocal to compute inverse propensity weights, $\hgammaipw(X_i)=1/\hat{e}(X_i)$.
Researchers typically estimate $e(x)$ via maximum likelihood, e.g., with a logistic regression model of $W$ on $X$, 
though other statistical machine learning approaches are used as well. 
The sample balance property \eqref{eq_mipw-sample-balance} implies that if the estimated propensity score $\hat{e}(x)$ converges to the true propensity score $e(x)$, 
$\hat{\mu}_1$ converges to $\mu_1$.

This approach suffers from two major drawbacks. 
The first is that the estimated weights $1 / \hat{e}(X_i)$ may not satisfy the sample balance property in Equation~\ref{eq_mipw-sample-balance}
with much (or any) accuracy, which can lead to bias. This can occur when the estimated weights are inaccurate,
 e.g., due to misspecification or a small sample size.
 In practice, it is often difficult to estimate conditional treatment probabilities, and 
inversion compounds this problem: small errors in estimating the propensity score can result in large errors
in estimating its inverse, especially when the propensity score is small.

The second drawback is that this approach can be very unstable. Observations with small estimated probabilities of treatment
tend to dominate the analysis. To some extent, this is inevitable, as we need to give rarely-treated observations 
large weights to ensure representativeness of the weighted sample. However, small errors in estimating treatment probability can lead to dramatic differences in the weights. If two units are assigned to treatment with probability $.05$, but we make small errors in 
estimating $.01$ for one and $.09$ for another, the first is nine times more influential than the second, though we err by only $.04$ in both cases.

To deal with these drawbacks, investigators resort to checks on the estimated inverse propensity weights $\hgammaipw(X_i)$.
For example, it is good practice to assess how well the balance property \eqref{eq_mipw-sample-balance} holds in the sample at hand by computing
the sample imbalance for some set of functions. 
Balance checks like this, for example comparing covariate means across groups, have a long history in observational causal inference \citep{cochran1965planning}. Comparing means is a natural approach 
when we think $m_1$ is approximately linear in $x$, as the maximal imbalance for a set of linear functions, e.g., $\model = \{ f(x)= \beta \cdot x : \norm{\beta}_2 \le 1 \}$, is a function of these marginal imbalances.
If imbalance is large, investigators typically start again from the beginning and fit a new propensity score model, 
repeating this process until they find one that achieves good balance.

\vspace{-.25cm}
\subsection{Balancing approach to weighting}
\label{sec:bal_weights_intro}

An alternative approach is to find weights $\hat{\gamma}$ by (approximately) solving for weights that satisfy the sample balance property \eqref{eq_mipw-sample-balance}, sometimes known as \emph{balancing weights}.
While different in its implementation, this is also an inverse propensity weighting estimator:
we can interpret balancing weights as a \emph{method of moments} estimate of the inverse propensity weights.
Specifically, as we discuss in Section \ref{sec:true_ipw}, the inverse propensity score weights $\gammaipw$ are the unique solution to the population balance moment conditions \eqref{eq_mipw-balance}.
The balancing approach finds 
weights $\hgamma$ that satisfy 
the corresponding sample balance moment conditions \eqref{eq_mipw-sample-balance},
i.e., find weights for which $\imbalance_{\model}(\hat\gamma)$ is small for a large enough model $\model$. 
%

Typically, this is done by solving an optimization problem to choose a weighting function $\hgamma(\cdot)$ to trade off imbalance  and some measure of the complexity of the weights, 
\begin{equation}
\label{eq_balancing-weights}
 \hgamma = \mathop{\argmin}_{\gamma}~~ \Bigg\{ \imbalance_{\model}^2(\gamma) \;\;+\;\; \frac{\sigma^2}{n^2} \sum_{W_i=1}\gamma(X_i)^2  \Bigg\},
\end{equation}
\noindent where $\sigma$ is a tuning parameter,
 or more generally, for increasing convex functions $\imbalancescale$ and $\dispersion$,
\begin{equation}
\label{eq_balancing-weights-general}
 \hgamma = \mathop{\argmin}_{\gamma}~~ \Bigg\{ \imbalancescale\{\imbalance_{\model}(\gamma)\} \;\;+\;\; \frac{1}{n^2} \sum_{W_i=1}\dispersion\{\gamma(X_i)\}  \Bigg\}.
\end{equation}
\noindent In the familiar special case that the weights control the \emph{maximal imbalance} for the linear model with $\ell_1$-bounded coefficients, $\model = \{\beta \cdot x  : \|\beta\|_1 \leq 1\}$,
the corresponding imbalance measure is the largest  imbalance in the individual elements of the covariate vector $x$:
\begin{equation} \label{eq:sbw}
\imbalance_{\model}(\gamma) = \max_{j=1,\ldots,p} \left| \frac{1}{n} \sum_{i=1}^n X_{ij} - \frac{1}{n}\sum_{i=1}^n W_i \gamma(X_i) X_{ij} \right| \text{ for }
\model = \{\beta \cdot x : \|\beta\|_1 \leq 1\}.
\end{equation}

Many existing approaches can be written as examples of Equation~\ref{eq_balancing-weights-general} with various 
choices of $\imbalancescale$ and $\dispersion$, including 
the \emph{stable balancing weights} approach \citep{zubizarreta2015stable};
the \emph{lasso minimum distance estimator} of the inverse propensity score \citep{chernozhukov2018automatic};
the \emph{regularized calibrated propensity score estimator} \citep{tan2020regularized}; and
\emph{entropy balancing} \citep{hainmueller2012balancing}.
In fact, even imputing $\mu_1$ by averaging the predictions of a least squares regression of $Y$ on $X$ fit on the treated units can be written in this form: it is a weighted average of observations $Y_i$ with
weights solving the problem above \eqref{eq_balancing-weights} for $\sigma^2=0$ \citep{chattopadhyay2021implied}. Similarly, ridge regression corresponds to $\sigma^2 > 0$ \citep{Ben-Michael2019}.

The special case \eqref{eq_balancing-weights} has a natural \emph{minimax} interpretation: the weights minimize the worst-case 
design-conditional mean squared error for $m_1 \in \model$ 
and $\Var[Y_i \mid X_i, W_i] \le \sigma^2$. 
Furthermore, when the model $\model$ is a convex set, they have an equivalent dual characterization
as a (non-parametric) least-squares estimate of the inverse propensity score. The weights are the function of the covariates, $\hat{\gamma}(X_i)$, that solves a penalized least squares problem,
\begin{equation}
  \label{eq:dual_l2_penalized_regression}
  \min_{\gamma} ~~~\underbrace{\frac{1}{n}\sum_{i=1}^n W_i \left(\gammaipw(X_i) - \gamma(X_i)\right)^2}_{\text{mean squared error}} \;-\; \underbrace{\left(\frac{1}{n}\sum_{i=1}^n\gamma(X_i) - \frac{1}{n}\sum_{i=1}^n W_i \gammaipw(X_i)\gamma(X_i) \right)}_{\text{mean zero ``noise''}} \;+\; \frac{\sigma^2}{n} \|\gamma\|_\calM^2,
\end{equation}
with penalty proportional to the square of the 
\emph{gauge} of the model, $\|\gamma\|_\model = \inf \{\alpha > 0 : \gamma \in \alpha \model\}$.\footnote{For example, for the linear model with $\ell_1$-bounded coefficients $\model = \{\beta \cdot x : \|\beta\|_1 \leq 1\}$ , the gauge is the $\ell_1$-norm of the coefficients 
of $\gamma(x) = \beta \cdot x$:  $\|\gamma\|_\model = \|\beta\|_1$,
and is infinite if 
$\gamma(x)$ is not linear in $x$.
}
The balancing property of the inverse propensity weights implies the middle term has mean zero, so up to this noise-like term, the minimax weights estimate $\gammaipw$ by minimizing penalized mean squared error.

The general case \eqref{eq_balancing-weights-general} has a qualitatively similar dual characterization, in which 
the inverse propensity weights $\gamma(X_i) = \eta_{\dispersion}(g(X_i))$
are parameterized in terms of an \emph{index function} $g(x)$, and the penalty is imposed on the index. For example, if the model were the set of linear functions with $\ell_1$-bounded coefficients $\{ \beta \cdot x : \norm{\beta}_1 \le 1\}$, this would be a generalized linear model $\gamma(X_i) = \eta_{\dispersion}(\beta \cdot X_i)$ with a penalty scaling with $\norm{\beta}_1$. In practice, $\dispersion$ is often chosen so that the \emph{link function} $\eta_{\dispersion}$ is the reciprocal of a logit or probit link, so when balancing we are fitting a familiar generalized linear model for the propensity score with a sparsity-inducing penalty. See Section~\ref{sec:dual} for details.

\section{Asymptotic behavior}
\label{sec:asymptotic}
%
%
%

We now give a brief overview of the asymptotic behavior of these weighting estimators to motivate our discussion in the next section. 
In essence, as long as imbalance in $\dm$ is asymptotically negligible relative to its standard error, $\hat{\mu}_1^{\aug}$ is asymptotically normal.

To be more precise, following \citet[see Corollary 6]{athey2018approximate}, when the
weights $\hgamma(X_i)$ are not too concentrated on any one individual and the noise $\varepsilon_i=Y_i-m_{W_i}(X_i)$ is not too heavy-tailed, 
negligible imbalance justifies the use of the standard Wald-type confidence interval $\hat \mu_1^{\aug} \pm z_{\alpha/2} \sqrt{\hat V}$, where $\hat V$
is a natural variance estimator and $z_{\alpha/2}$ is a quantile of the standard normal distribution.
We state a result for estimating 
$\tilde \mu_1 = \frac{1}{n}\sum_{i=1}^n m_1(X_i)$.\footnote{By focusing on  accuracy in estimating $\tilde \mu_1$, a sample analog of $\mu_1$, we drop the sampling variation term from the error decomposition in Equation~\ref{eq:weights_error_aug}. In a sense, because all we 
know about the distribution of $X_i$ is summarized by the sample $X_1 \ldots X_n$, it is inevitable that every estimator of $\mu_1$ is 
an estimate of $\tilde \mu_1$ as well, and a more precise one at that. See \citet{imbens2004nonparametric}.}
\begin{proposition}
 \label{proposition:clt}
Let $\hat m_1^{(-i)}$ be a consistent cross-fit estimator of $m_1$
and $\Gamma :=  \sqrt{n^{-1}\sum_{i=1}^n W_i \hgamma(X_i)^2}$ be the root-mean-squared weight. If imbalance in $\dm$
(see Equation~\ref{eq:weights_error_aug}) is $o_p(n^{-1/2} \Gamma)$, then
 \[  \frac{\hat\mu^{\aug} -  \tilde \mu_1}{\sqrt{\hat V}} \to N(0,1) \quad \text{ where } 
 \quad \hat V = \frac{1}{n^2}\sum_{i=1}^n W_i \hgamma(X_i)^2 \{Y_i - \hat m_1^{(-i)}(X_i)\}^2 \] 
 if, in addition, each weight's square is small relative to the sum of them all ($n\Gamma^2$), and the noise sequence $\varepsilon_1 \ldots \varepsilon_n$ is uniformly square integrable conditional on the design.
\end{proposition}
\noindent In Appendix~\ref{sec:appendix-asymptotic-normality} we include a more formal statement and a simple proof. 
In essence, because the noise term dominates the error decomposition \eqref{eq:weights_error_aug} and is 
a sum of conditionally independent mean-zero random variables, 
the estimate satisfies a central limit theorem. 
We will from this point on focus on the condition that imbalance be $o_p(n^{-1/2})$.  If there is sufficient overlap (Assumption~\ref{a:regular-overlap}), this will tend to be equivalent to the condition that imbalance be $o_p(n^{-1/2} \Gamma)$ stated in Proposition~\ref{proposition:clt}, as the root-mean-squared weight $\Gamma$ will tend to be of constant order; we will get more extreme weights 
when overlap is poor, and in those circumstances are allowed a bit more imbalance.


Furthermore, if we assume sufficient overlap (Assumption \ref{a:regular-overlap}), then a clear story of optimality emerges: all \emph{regular} estimators 
are asymptotically equivalent (see, e.g., van der Vaart, \citeyear{van2000asymptotic}, Chapter 25). 
In particular, they are equivalent to $\hat\mu_1^{aug}$ with weights $\hgamma=\gammaipw$ and a consistent estimator $\hat m_1$ of the outcome regression $m_1$. 
We get a regular estimator when we achieve negligible imbalance using weights that balance a sufficiently rich model $\model$: one that is nonparametric in the sense that $\vectorspan(\model)$ contains an approximation to every square integrable function.
If instead we use weights that balance a less rich model, similar results hold, albeit with a weakened sense of regularity. 
When we do this, the balancing weights
\eqref{eq_balancing-weights} converge to a \emph{projection} of $\gammaipw$ 
onto the span of the functions in the model $\model$ rather than $\gammaipw$ itself,
resulting in an estimator that has lower variance but is substantially less robust \citep[Remark 3]{hirshberg2018augmented}. 


\section{Balance measures and models}
\label{sec:balance_models}

In the balancing approach of Section~\ref{sec:bal_weights_intro},
we attempt to control imbalance in $\dm$ by
controlling the maximal imbalance 
for candidates in a model $\model$.
In this section, we discuss possible choices for this model and connections to balance criteria in the literature.
We first consider the simple case that we have only binary covariates before moving on to the general case with continuous covariates.


\vspace{-.25cm}
\subsection{Balancing binary covariates}
\label{sec:balancing-binary}

\subsubsection{Models with no interactions}
\label{sec_balancing_marginally}

We begin with the case where we have $p$ binary covariates $X_1,\ldots,X_p \in \{0,1\}$, for example indicating employment status or high school education.
In this case, we consider the common practice of only balancing covariates marginally,
which is equivalent to balancing a linear model without interactions.
This approach is implicit in typical balance checks in observational studies that compare the fraction of units in the treated group with each attribute to the corresponding fraction in the population. Written explicitly, the imbalance term is
\begin{equation}
    \label{eq:imbalance_margin_cat}
    \begin{aligned}
   \imbalance^2_{\model}(\gamma) &= \sum_{j = 1}^p \abs*{\frac{1}{n}\sum_{i=1}^n X_{ij} - \frac{1}{n}\sum_{i=1}^n W_i\gamma(X_i) X_{ij} }^2 \quad \text{ for }\quad \\
   \model &= \set*{ m(x) = \beta \cdot x : \norm{\beta}_2 \le 1 }.
   \end{aligned}
\end{equation}
Finding weights to balance marginal proportions has a long history in survey statistics, where it is known as \emph{raking} \citep[e.g.,][]{Deming1940, deville1992calibration}. 
In causal inference such balance measures are commonly used to assess covariate balance across treatment groups \citep[e.g.,][]{cochran1965planning, rosenbaum1985constructing}.

If the conditional mean outcome $m_1$ belongs to this model $\model$, then the estimator's design-conditional bias is
\begin{equation}
  \label{eq:bias_linear}
  \beta \cdot \left(\frac{1}{n}\sum_{i=1}^n X_i - \frac{1}{n}\sum_{i=1}^n W_i\hgamma(X_i) X_i\right).
\end{equation}
When $\norm{\beta}_2 \le 1$, this is bounded by $\imbalance_\model(\gamma)$ from Equation~\ref{eq:imbalance_margin_cat}. In particular, weights that exactly balance each covariate marginally will produce an unbiased estimator. 

However, we seldom believe that $m_1$ is linear, so balancing the margins alone is insufficient to remove bias.
To see this, consider the decomposition of 
$m_1$ into its best linear approximation $\tilde m_1(x) := \tilde\beta \cdot x$ and the residuals 
$r(x) := m_1(x) - \tilde m_1(x)$.
Balancing margins leaves additional \emph{approximation error} based on the difference between the approximation $\tilde m_1$ and $m_1$ itself. Generally, the design-conditional bias is
\begin{equation}
  \label{eq:bias_basis}
  \tilde \beta \cdot \underbrace{\left(\frac{1}{n}\sum_{i=1}^n X_i - \frac{1}{n}\sum_{i=1}^n W_i\hgamma(X_i) X_i\right)}_{\text{marginal imbalance}} \;\;+\;\; \underbrace{\left(\frac{1}{n}\sum_{i=1}^n r(X_i) - \frac{1}{n}\sum_{i=1}^n W_i\hgamma(X_i) r(X_i)\right)}_{\text{imbalance in the approximation error}}.
\end{equation}
When we use balancing weights like those discussed above 
for a linear model \eqref{eq_balancing-weights-general}, we make no attempt to balance approximation error.
We can improve approximation error by instead balancing a \emph{basis expansion} $\phi(x)$ that includes interactions; however, this leaves us with two competing goals. 
We would like to balance an expressive enough model to give a good approximation of $m_1(x)$ and also like to ensure that we are able to get reasonable balance for that model.

\subsubsection{Models with strong interactions} 
\label{sec_balancing_all}

Balancing models with interactions is often sensible.
For example, we might believe that someone will respond differently if they are both employed \emph{and} a high school graduate, so we would want to balance the interaction between the two.
More formally, for our vector of $p$ binary covariates $X_{i} = (X_{i1}, \ldots, X_{ip})$ and multi-index vector $j = (j_1, ..., j_K) \in \{0,1\}^p$, 
let $X_{i}^{j}$ be the covariate interaction $X_{i1}^{j_1} \cdot \ldots \cdot X_{ip}^{j_p}$, so the interaction of being employed and a high school graduate
corresponds to an index vector $j$ with ones in those two coordinates, say with $j_1=j_2=1$, and zeros elsewhere.

We can attempt to balance \emph{all} possible higher-order interactions. In this notation, letting sums over $j$ range over $\{0,1\}^p$ implicitly, the imbalance term is 
\begin{equation}
    \label{eq:imbalance_interact}
    \begin{aligned}
   \imbalance_{\model}^2(\gamma) &=  \sum_{j} \abs*{ 
   \frac{1}{n} \sum_{i=1}^n X_{i}^{j} - \frac{1}{n}\sum_{i=1}^n W_i \gamma(X_i) X_{i}^{j}
   }^2 \quad \text{ for } \quad \\ \model &= \set{ m(x) = \sum_{j} \beta_j x^j : \sum_{j} \beta_j^2 \le 1}.
   \end{aligned}
\end{equation}

For low-dimensional covariates, this type of balance measure is commonly used in sample surveys and causal inference. 
If we can find weights such that the imbalance in Equation~\ref{eq:imbalance_interact} is zero, then these weights will completely remove the bias without any
assumptions on the conditional mean $m_1$.
In other words, when we have categorical covariates, we can remove bias for every possible conditional mean function $m_1$ by finding weights that \emph{exactly} balance every interaction --- so long as such weights exist.

\subsubsection{Models with weak interactions}

The challenge with balancing all higher-order interactions is that the set of interactions to balance grows exponentially in the number of covariates, making it infeasible to balance everything well. 
Thus, the curse of dimensionality often prevents us from exactly balancing all higher-order interactions. 
Even if we allow for approximate balance,
the imbalance measure in Equation~\ref{eq:imbalance_interact} gives the same importance to the marginal imbalance in a covariate and the imbalance in the $p^{th}$ order interaction $X_{i1} \cdot X_{i2} \cdot ... \cdot X_{ip}$. 
This is appropriate if we think the effects of high-order interactions may be as large as the main effects, but models in which effects tend to be smaller for higher order interactions, 
like the one
used in \citet[Equation 9]{benmichael2021_multical},
tend to be more plausible.

To encode the classical assumption that the coefficients corresponding to main effects
are larger than the effects of first-order interactions, which are larger than those for second-order interactions, and so on, 
we define scale factors $\lambda_j$ as a decreasing function of the interaction order $\norm{j}_1$. We can then  control the maximal imbalance over a model in which the $\lambda_j^{-1}$-scaled effects are bounded.
\begin{equation}
    \label{eq:binary-rkhs} 
\begin{aligned}
\imbalance^2_{\model}(\gamma) &= \sum_{j} \lambda_j^2 \left|\frac{1}{n}\sum_{i=1}^n X_i^j - \frac{1}{n}\sum_{i=1}^nW_i \gamma(X_i) X_i^j\right|^2 
\quad \text{for} \quad \\
\model &= \set*{m(x) = \sum_j \beta_{j} x^j : \sum_j \beta_j^2 \ /\ \lambda_j^2 \le 1}.
\end{aligned}
\end{equation}
 

Taking $\lambda_j=0$ corresponds to assuming away the presence of an effect of $x^j$, so we do not penalize imbalance in the corresponding interaction, whereas taking $\lambda_j=\infty$ corresponds to assuming  no limits on the effect, so we must balance the corresponding interaction
perfectly. 
In Section~\ref{sec_balancing_marginally} we discussed the case in which $\lambda_j$ is nonzero only for main effects.
One common approach is to adjust exactly for main effects and low-order interactions and not at all for higher-order ones.
Sometimes exact balance on low-order terms is combined with approximate balance on higher order ones by taking $\lambda_j=\infty$ only for small $\norm{j}_1$;
this is called \emph{fine balance} in the literature on matching in causal inference \citep{rosenbaum2007minimum}. 

\subsection{Balancing continuous covariates}
\label{sec:balancing-continuous-covariates}


\subsubsection{The basis expansion perspective}

The key change when we work with continuous covariates is that interactions are not the only source of complexity: functions of even one continuous covariate can be complex and nonlinear. In essence, we face the curse of dimensionality even in one dimension, as non-parametric models require an \emph{infinite} basis. For example, the natural generalization of the previous model \eqref{eq:binary-rkhs} is a Taylor series model, in which we sum over all powers of each covariate ($j \in \mathbb{N}^p$), instead of the binary ones ($j \in \{0,1\}^p$). In this model we penalize higher order terms, e.g., $x_1 \cdot \ldots \cdot x_4$ or $x_1^4$, irrespective of whether they are interactions or powers of one covariate. 
More generally, we can use a similar model specified in terms of any set of basis functions $\phi_1(x),\phi_2(x),\ldots$ and corresponding scale factors $\lambda_1,\lambda_2,\ldots$.
\begin{equation}
    \label{eq:imbalance_phi_infinite}
\begin{aligned}
    \imbalance^2_{\calM}(\gamma) &= \sum_j \lambda_j^2 \abs*{\frac{1}{n}\sum_{i=1}^n \phi_j(X_i) - \frac{1}{n}\sum_{i=1}^n W_i \gamma(X_i) \phi_j(X_i)}^2 \qquad \text{ for } \\
    \model &= \set*{m(x) = \sum_j \beta_{j} \phi_j(x) : \sum_j \beta_j^2 \ / \ \lambda_j^2 \le 1}.
\end{aligned}
\end{equation}
To better isolate first order effects from higher order ones in a sequential manner, it is common to use a basis of \emph{orthogonal polynomials}
instead of the Taylor series-inspired basis above; for example, the Hermite polynomials $1,x_1,x_2,x_1^2-1,x_2^2-1,x_1x_2$, written here for two covariates.
This isolation can make specifying  $\lambda_j$ a bit more intuitive. 
Another option, motivated by Fourier series, is to use an orthogonal basis of sines and cosines of different frequencies. We cannot balance sines and cosines of all frequencies accurately just like we cannot balance all polynomials, so we use models that restrict or rule out effects associated with high frequency terms.

As before, taking $\lambda_j=0$ corresponds to assuming away any effect for $\phi_j(x)$, so  $\imbalance_\model(\gamma)$ is unaffected by imbalance in $\phi_j$, and taking $\lambda_j=\infty$ corresponds to assuming no limits on that effect and therefore requires exact balance in $\phi_j$. 

\subsubsection{The function perspective}

While the basis expansion perspective can be useful when thinking about the balancing problem, specifying an infinite-dimensional model $\model$
directly in terms of basis functions and scale factors ($\phi_j, \lambda_j$) is hard to do well. We can also view things from another perspective,
focusing on the properties of the functions rather than their basis expansion.
The theory of reproducing kernel Hilbert spaces (RKHSes) provides one approach for doing so. Subject to some constraints, the model $\model$ in Equation~\ref{eq:imbalance_phi_infinite} is the unit ball $\{m(x) : \norm{m}_\model \le 1\}$ of an RKHS,
and furthermore, the unit ball of every RKHS can be described like this  \citep[see, e.g.,][Chapters 2 and 4]{cucker2007learning}.\footnote{Specifically, these constraints are that the basis $\phi_j$ is orthonormal with respect to some probability measure and the decay factors $\lambda_j$ have a finite sum.}  

One of the most common ways to characterize functions is their smoothness. For this, it is natural to work with Sobolev spaces: RKHSes defined in terms of averages of a function's derivatives. We can characterize the unit ball of a \emph{Sobolev norm} in terms of the coefficients of Fourier series expansions. For example, one Sobolev norm for functions on $[0,\pi]^2$ is
\begin{align}
     \norm{f}^2 
&= 
     \frac{1}{\pi^{2}} \int_{[0,\pi]^2} \ \abs*{f(x_1,x_2)}^2 + \abs*{\frac{\partial }{\partial x_1}f(x_1,x_2)}^2 +
     \abs*{\frac{\partial }{\partial x_2}f(x_1,x_2)}^2 +
     \abs*{\frac{\partial^2 }{\partial x_1 x_2}f(x_1,x_2)}^2 \label{eq:sobolev-derivative} \\
 &= \sum_{j \in \mathbb{Z}^2} \beta_j^2\ /\ \lambda_j^2 \text{ with } 1/\lambda_j^{2} = (1+j_1^2)(1+j_2^2) \text{ for } f(x)=\sum_{j \in \mathbb{Z}^2} \beta_j  \cos(j_1x_1 + j_2x_2) \label{eq:sobolev-fourier}.
 \end{align}
In these models, our assumptions are encoded in the derivatives included in the norm.
If we expect $m_1$ to be relatively smooth, including higher order derivatives encodes this assumption. 
If we believe that most of the complexity lies along the coordinate axes, i.e., that a function tends to vary more with one covariate than with a mixture of a few, including higher order mixed partial derivatives (as in \ref{eq:sobolev-derivative}) encodes this assumption.  If instead of the cube $[0,\pi]^2$, we define our space on all of $\R^2$ by averaging our derivatives over the normal distribution, we get an RKHS with a norm characterized in terms of a Hermite polynomial rather than Fourier basis.\footnote{In particular, if  we replace the average over the uniform distribution on the square in 
 \eqref{eq:sobolev-derivative}
 with an average over the standard bivariate normal, we get a norm equivalent to a variant of \eqref{eq:sobolev-fourier} in 
 which the $f_k$ are the coefficients in a a series expansion \smash{$f(x) = \sum_{j \in \mathbb{Z}_+^2} f_j h_j(x)$} where $h_j$ is the bivariate Hermite polynomial of order $j$ \citep[see e.g.,][]{bongioanni2006sobolev}.} We have talked about functions on $\R^2$ only to simplify notation: All these concepts readily extend to higher dimensions and are discussed generally in our references. We discuss them at greater length in Appendix~\ref{sec:more-on-models}.
 
 
Balancing weights for Sobolev space models and RKHS models more generally are discussed in \citet{Hazlett2019}, \citet{hirshberg2019minimax}, \citet{kallus2020generalized}, \citet{singh2020reproducing}, and \citet{wong2018kernel}. RKHS models are convenient because optimization problems involving infinite series (as in \ref{eq:imbalance_phi_infinite}) can be reduced to finite dimensional problems via the so-called \emph{kernel trick}. See Appendix~\ref{sec:RKHS-computation} for details.

So far we have focused on models with relatively weak interaction effects. These have the benefit of being fairly small, as qualitatively there are more ways for a function to vary in multiple directions than in one. We could consider a larger model, 
for example the set of \emph{Lipschitz functions} $\model=\set{m : \abs{m(x)-m(x')} \le \norm{x-x'} \text{ for all } x,x'}$. This model, however, can be much too large to achieve negligible bias with the associated IPW estimator. As a matter of fact, the IPW estimator with the minimax balancing weights \eqref{eq_balancing-weights} for this model reduces, for small enough noise level parameter $\sigma$, to the matched difference in means after nearest neighbor matching \citep{armstrong2018finite,kallus2020generalized}. This is known to have non-negligible bias for $p \ge 4$ covariates \citep[Theorem 1]{abadie2006large}.

\section{Ensuring good enough balance}
\label{sec:trade-offs}

The discussion above focuses on balancing richer and more complex models.
But there is an inherent trade-off: the larger the model, the harder it is to balance everything in it. Achieving negligible imbalance in the regression error (in the sense of Section~\ref{sec:asymptotic}),
is essential for straightforward inference.\footnote{Although see \citet{armstrong2018finite} for a discussion of bias-aware confidence intervals.}
We now turn to understanding when this happens.


\subsection{Comparing to the true inverse propensity weights}
\label{sec:plugin}

We first focus on the unaugmented estimator, so $\delta m = -m_1$,
and assume our model is valid in the sense that $-m_1 \in \model$; thus,
imbalance in $m_1$ is bounded by $\imbalance_{\model}(\hgamma)$. We can bound this by comparing it to $\imbalance_\model(\gammaipw)$, the maximal imbalance achieved by the  true inverse propensity weights. For example, because the minimax weights $\hgamma$ minimize the objective function in Equation~\ref{eq_balancing-weights},
\begin{equation}
\label{eq_balancing-weights-comparison}
\imbalance_{\model}^2(\hgamma) \;\;+\;\; \frac{\sigma^2}{n^2} \sum_{W_i=1}\hgamma(X_i)^2 \le
\imbalance_{\model}^2(\gammaipw) \;\;+\;\; \frac{\sigma^2}{n^2} \sum_{W_i=1}\gammaipw(X_i)^2.
\end{equation}
As \citet{hirshberg2018augmented} discuss, because the estimated weights $\hgamma$ converge to the true inverse propensity weights $\gammaipw$, the difference between the variance terms is asymptotically negligible. As a result, this comparison implies that up to a negligible term, $\imbalance_{\model}(\hgamma)$ is less than  $\imbalance_\model(\gammaipw)$, which is relatively easy to characterize. Because the inverse propensity weights $\gammaipw$ have the population balance property \eqref{eq_mipw-balance}, 
$\imbalance_{\model}(\gammaipw)$ is the maximum (over all elements of $\calM$) of an average of independent and identically distributed terms with mean zero. Hence, we can use the central limit theorem to make our argument.

Consider the model $\{\beta \cdot \phi(x) : \|\beta\|_1 \leq 1\}$ for a basis expansion $\phi(x) \in \R^p$. By H\"older's inequality, the corresponding maximal imbalance is the imbalance in the least-balanced basis function:
\begin{equation}
  \label{eq:ipw_linf_imbalance}
  \begin{aligned}
  \imbalance_\calM(\gammaipw) &= \max_{j=1,\ldots,p} \left| \frac{1}{n}\sum_{i=1}^n \phi_j(X_i) -  W_i \gammaipw(X_i) \phi_j(X_i)\right| \quad \text{ for } \quad \\
  \model &= \set*{\beta \cdot \phi(x) : \|\beta\|_1 \leq 1}.
  \end{aligned}
\end{equation}
We know, from the central limit theorem, that the imbalance in the basis function $\phi_j$ is approximately normal with standard deviation on the order of $n^{-1/2}$ in large samples. 
In essence,\footnotemark\ this maximal imbalance \eqref{eq:ipw_linf_imbalance} is the maximum of the individual imbalances in $p$ basis functions and will therefore behave like the maximum of $p$ 
mean-zero normal random variables with standard deviation on the order of $n^{-1/2}$; that is, it will be $O_p\left(\sqrt{\log(p)/{n}}\right)$, 
depending very weakly on the number of basis functions $p$.
\footnotetext{The is imprecise only in that it ignores the presence of the absolute value. Because it is there, we can interpret the maximum as one over imbalances in $2p$ basis functions: $\phi_j$ and $-\phi_j$ for all $j$.}
A \emph{symmetrization} argument allows us to formalize this intuitive argument 
without appealing to asymptotics \citep[e.g.,][Lemma 6.4.2]{vershynin2018high}. \citet{athey2018approximate}
use this approach to analyze balancing weights for high dimensional linear models.

We can use similar arguments for other models, including infinite-dimensional ones.
In general, the ``size'' of $\calM$ is determined to the number of distinct functions we would need to have an approximation,
up to a certain error tolerance, of  every element of $\model$. 
Using a \emph{chaining argument} \citep[e.g.,][Chapter 8]{vershynin2018high} that combines approximations at different error tolerances with bounds for finite sets, 
we can show that $\imbalance_{\calM}(\gammaipw) = O_p(n^{-1/2})$ for many models: essentially, those called \emph{Donsker} classes.\footnote{For such models, the maximum $\imbalance_{\model}(\gammaipw)$ of the individual imbalances in the functions it contains
is with high probability barely larger than we would expect any individual imbalances to be because the latter are \emph{highly correlated};
the maximum of $p$ normals as above is largest, and on the order of $\sqrt{\log( p)/n}$, when they are independent.}$^,$\footnote{To be more precise,
this holds when both $\calM$ and the related set $\{f(w,x) = w\gammaipw(x)m(x) : m \in \calM\}$ are Donsker. If the \emph{strict overlap} condition that $\gammaipw$ is bounded holds, this is true if $\calM$ itself is Donsker, but it can be stronger otherwise. See, e.g., 
\citet[Remark 4]{hirshberg2018augmented}.}
This holds, for example, if $\calM$ is the unit ball of a Sobolev space of functions on $\R^{p}$ 
with bounded derivatives of order $s > p/2$ \citep{nickl2007bracketing}. 
\citet{kallus2020generalized} and \citet{wong2018kernel} use these bounds to characterize the behavior of balancing estimators that work with infinite-dimensional models.

Unfortunately, this approach is not sufficient to show that the imbalance in the conditional mean outcome $m_1$ is $o_p(n^{-1/2})$.  To see this, consider the case of a model $\model$ that contains only a single function: the conditional mean $m_1$. In this case,
\begin{equation}
  \label{eq:single_function_imbalance}
  \imbalance_\calM(\gammaipw) = \left|\frac{1}{n}\sum_{i=1}^n m_1(X_i) - W_i \gammaipw(X_i) m_1(X_i)\right| \quad\quad \text{ for } \quad\quad \calM = \{m_1\}.
\end{equation}
By the central limit theorem,
we know that $\sqrt{n} \imbalance_\calM(\gammaipw)$ will not converge to zero but will instead converge to the absolute value of a normally-distributed random variable 
with mean zero and variance $\var[\{1 - W_i \gammaipw(X_i) \} m_1(X_i)]$.\footnotemark\  Therefore, even in this simple case, the analytical approach of bounding the imbalance achieved by the estimated weights by that achieved by the true inverse propensity weights cannot establish negligible imbalance.

\footnotetext{The IPW estimator with the true inverse propensity weights $\gammaipw$ is asymptotically unbiased and normal, as the imbalance term is perfectly centered: we get additional variance, but not bias, from our non-negligible imbalance. However, this delicate centering property is not necessarily preserved even by estimated weights that are quite close to the true ones in terms of, e.g., mean squared error.}

\subsection{The role of augmentation}
\label{sec:role-of-augmentation}
We can do better with this approach when analyzing the augmented estimator we discuss in Section \ref{sec:aipw}.
The key idea is that rather than analyzing balance in a \emph{fixed} model, by augmenting the estimator we can restrict our attention to a \emph{shrinking} sequence of models. We can show that the imbalance achieved by the inverse propensity weights on this sequence of models is indeed negligible.

For intuition, consider again the case where the model has a single element, 
which we now index by the sample size:
$\calM_n = \{\delta m_n\}$. This represents the imbalance for the augmented estimator: the regression error $\dm_n = \hat m_{1,n} - m_1$ will shrink as we have more observations. It is convenient to think of $\hat m_1$ as cross-fit, i.e., estimated
on a sample other than $\{ (W_i,X_i,Y_i) : i \le n\}$. This allows us to think of $\dm$  as \emph{non-random} in the sense
that it does not depend on this sample.\footnotemark\ Now the imbalance depends on the sample size: \footnotetext{This is not necessary for all arguments, e.g., it is not required by those of \citet{hirshberg2018augmented}.
However, it is increasingly common practice, has been shown to be help in some simulation studies, and
can be done without loss of efficiency asymptotically by averaging estimators that fit $\hat m_1$ on different subsamples
\citep[see, e.g.][]{chernozhukov2018double, schick1986asymptotically, zheng2011cross}.}
\[
  \imbalance_{\calM_n}(\gammaipw) = \left|\frac{1}{n}\sum_{i=1}^n \delta m_n(X_i) - W_i \gammaipw(X_i) \delta m_n(X_i)\right| \quad \text{ for } \quad \model_n = \set{ \delta m_n}.
\]
By the same argument we used for $\model=\set{m_1}$, if $\Var[\{1-W_i\gammaipw(X_i)\}\delta m_n(X)]$ converges to zero, then $\sqrt{n} \imbalance_{\calM_n}(\gammaipw)$ will converge to zero in probability and the estimator will achieve asymptotically negligible imbalance.
Given mild assumptions, this quantity converges to zero if $\hat m_1$
is a consistent estimator.\footnotemark

\footnotetext{For example, sufficient overlap (Assumption~\ref{a:regular-overlap}) and boundedness of $\norm{\dm_n}_{\infty}$ suffice \citep[see Proof of Theorem 1]{hirshberg2018augmented}.}

To view this phenomenon more generally, suppose we have a valid model $\model_n$ with many elements, all of them are shrinking like $\dm_n$ in the previous example.
In particular, suppose that we begin with some a-priori model $\model_0$ containing $\dm_n$ for all $n$, and we know the rate of convergence of our estimator $\hat m_{1,n}$,
so at each time-step we are able to rule out that possibility that our regression error $\norm{\dm_n}_{L_2(P)}$ exceeds some bound $r_n$
that shrinks to zero.\footnote{In discussion of the role of augmentation within the balancing tradition,
it has been common to focus on consistency in the strong sense that $\norm{\dm}_{\model} \to 0$ rather than in the mean-square sense $\norm{\dm}_{L_2(P)} \to 0$ that we use here. While simpler, that approach has some disadvantages. See Appendix~\ref{sec:strong-norm-consistency} for a discussion.} Incorporating this knowledge leaves us with the smaller model $\model_n = \set{m \in \model_0 : \norm{m}_{L_2(P)} \le r_n}$.
It is reasonable to ask whether this is enough that the maximal imbalance would still satisfy $\sqrt{n}\imbalance_{\model_n}(\gammaipw) \to 0$.
The answer depends on the size the a-priori model $\model_0$, but essentially a sufficient condition on $\model_0$ is that $\imbalance_{\model_0}(\gammaipw) = O_p(n^{-1/2})$.\footnotemark\ 

\footnotetext{To be precise, it is yes when $\calM_0$ and $\{f(w,x) = w\gammaipw(x)m(x) : m \in \calM_0\}$ are Donsker
and the maximal variance of a term for an individual element of $\model_n$, $\max_{\dm \in \model_n}\Var[(1-\gammaipw(X_i)W_i)\dm(X)]$,
shrinks to zero. Generalizing the one-element case discussed above, 
sufficient overlap (Assumption~\ref{a:regular-overlap}) and boundedness of $\max_{m \in \model_0}\norm{m}_{\infty}$ suffice for the latter.
A slight generalization of this case is addressed in \citep[see Proof of Theorem 1]{hirshberg2018augmented}.
This phenomenon, the inheritance of convergence of $\sqrt{n}\imbalance_{\calM_n}(\gammaipw)$ 
from the convergence of the largest variance of a term $(1-\gammaipw(X_i)W_i)\dm(X)$ in the averages
we are maximizing over, is called the asymptotic equicontinuity of the empirical process,
and is the defining feature of Donsker classes \citep[e.g.,][Section 2.1.2]{van1996weak}.}

From this, it would appear that we need to know the valid a-priori model $\model_0$ \emph{and} the rate of convergence of our estimator $\hat{m}_{1,n}$.
However, 
\citet[see Section 2]{hirshberg2018augmented}
show that the minimax weights \eqref{eq_balancing-weights} for an a-priori model $\model_0$ of appropriate size achieve negligible imbalance over the shrinking model sequence $\model_n = \{ \dm \in \model_0 : \norm{\dm}_{L_2(P)} \le r_n\}$ for any rate of convergence $r_n$ tending to zero. Thus, all you need is a valid a-priori model for $\dm$, i.e., one satisfying $\dm \in \model_0$. 
If we have a valid model for $m_1$ itself, it is easy to obtain something close enough. We can take $\model_0$ to be the
convex hull of $\hat m_1 - \model' :=\{\hat m_1 - m : m \in \model'\}$, which will contain $\dm$  when $m_1 \in \model'$
\citep[Section 2.6]{hirshberg2018augmented}.

\subsection{Other perspectives}
\label{sec:beyond-the-plug-in}

While reducing the analysis to characterizing the maximal imbalance achieved by the true inverse propensity weights
can be informative, it has limits, and it can be helpful to consider different perspectives.
To guide this discussion, note that we can expand the conditional bias term in Equation~\ref{eq:weights_error_aug} into two terms,
\begin{equation}
    \label{eq:aug_bias_dr}
    \frac{1}{n}\sum_{i=1}^nW_i \delta\gamma(X_i) \dm(X_i) \;+\; \frac{1}{n}\sum_{i=1}^n \left\{W_i \gammaipw(X_i) - 1\right\}\dm(X_i) \quad \text{ where } \quad  \delta\gamma = \hat{\gamma}  - \gammaipw.
\end{equation}
The second term in Equation~\ref{eq:aug_bias_dr} is the imbalance in the error function $\dm$ with the true 
inverse propensity weights $\gammaipw$; as we discuss above, this will be $o_p(n^{-1/2})$ if $\hat m_1$ is consistent. 
Therefore, we can focus our attention on the first term, the inner product between the error in estimating the true inverse propensity weights, $\dgamma$, and the error in estimating the outcome regression, $\dm$. The conditional bias is asymptotically negligible \emph{if and only if} this inner product is.

We can use the Cauchy-Schwartz inequality to bound this inner product in terms of the mean squared error of the estimated inverse propensity weights, $\hgamma$, 
and the estimated outcome regression, $\hat{m}_1$.
This gives a simple sufficient condition 
for the conditional bias  to be asymptotically
negligible: the product of the rates of convergence of $\hat\gamma$ and $\hat m_1$, $\norm{\delta\gamma}_{L_2(\Pn)}\norm{\dm}_{L_2(\Pn)}$, 
must be $o_p(n^{-1/2})$. 
This argument is widely used, and results in the concept of \emph{double robustness}: that we need not estimate one with any degree of accuracy, 
so long we estimate the other well enough to compensate. 

However, this argument also has limits.
For instance, if the two rates of convergence are the same, they must both be faster than $n^{-1/4}$.
In contrast, the argument in Section~\ref{sec:role-of-augmentation} essentially
requires $m_1$ to be estimable at a $n^{-1/4}$ rate but places no assumptions on $\gammaipw$. 
In essence, the two arguments use different ways of bounding the inner product: the double-robustness argument uses bounds on the \emph{magnitudes} of the residuals $\delta\gamma$ and $\dm$, while the balancing argument instead shows that they approximately \emph{orthogonal} \citep[Section 1.4]{hirshberg2018augmented}.

Because these heuristics can be pessimistic, it is important to keep their shortcomings in mind. 
There are estimators that have this same asymptotic behavior if \emph{either} $\gammaipw$ or $m_1$ is smooth enough to be estimable at faster than
$n^{-1/4}$ rate, as well as in the range between these extremes in which both are to some degree smooth but neither estimable at $n^{-1/4}$ rate;
these asymptotics are called \emph{optimally double robust} because it is known that these requirements cannot be improved \citep{mukherjee2017semiparametric, robins2009semiparametric}. 
While the estimators known to exhibit optimal double robustness are relatively complex,
a careful analysis of the AIPW estimator can show it is not far off. In particular,
\citet{newey2018cross} show that with specialized estimators of both $\gammaipw$ and $m_1$ that are amenable to careful
analysis of the orthogonality of $\delta\gamma$ and $\dm$, the AIPW estimator
has nearly the same asymptotic behavior.\footnotemark\ Work like this, which brings us closer to understanding what is needed for straightforward inference than arguments based on maximal imbalance or the error rate product, is an important area of ongoing research \citep[e.g.][]{kennedy2020optimal}.
\section{Practical considerations}
\label{sec:misfit_toys}

\subsection{Translation invariance}
\label{sec:translation-invariance}

The average treatment effect $\tau = \mu_1 - \mu_0$ is translation invariant; it is not affected by the way the outcomes are centered. It aids our intuition to use an estimator $\hat\tau = \hat \mu_1 - \hat \mu_0$ with the same property: if we estimate $\mu_1$ when $m_1(x)=f(x)$, then we should estimate $\mu_1 + t$ when $m_1(x)=f(x) + t$. This invariance is imposed by our imbalance measure when our model includes all translations of every function in it.
  \[
   \begin{aligned}
   &\imbalance_{\model}(\gamma) = 
     \begin{cases}
       \imbalance_{\model_0}(\gamma) & \text{ if }\ \frac{1}{n}\sum_{i=1}^n W_i \gamma(X_i) = 1\\
       \infty &  \text{ if }\ \frac{1}{n}\sum_{i=1}^n W_i \gamma(X_i) \neq 1
   \end{cases} 
   \quad\ \text{ for } \\ &\quad \model = \set*{ m(x) + t : m \in \model_0,\  t \in \R}.
  \end{aligned}
   \]
This is equivalent to imposing the constraint that our weights average to one. For model based weights, it's common practice to ensure translation invariance by normalizing them to average to one, using $\hat\gamma(X_i) = \hat e(X_i)^{-1} / \{ n^{-1} \sum_{i=1}^n W_i \hat e(X_i)^{-1}\}$ instead of $\hat \gamma(X_i)=\hat e(X_i)^{-1}$.

\subsection{Sample boundedness}
\label{sec:sample-boundedness}

Constraining the weights to be non-negative and average to one disallows \emph{extrapolation} outside of the convex hull of the data. Intuitively, using weights that interpolate the data limits the extent of biases due to misspecification of the model $\mathcal{M}$ relative to what is possible with extrapolation. In particular, this ensures that the estimate is sample bounded, i.e. that it lies between the minimum and maximum observed values of the outcome. 
This rules out nonsensical estimates when outcomes are known to be bounded, such as for binary and survival outcomes. 
However, because balancing weights for rich models converge to the inverse propensity weights, and they exceed one, we tend to get negative weights when we are balancing a small --- and likely to be misspecified --- model. 
When this is the case, using a richer model tends to be a better fix than an explicit non-negativity constraint or, at least, a good complement to one.

\subsection{The bias-variance trade-off}
\label{sec:bias-variance-tradeoff}

The usual interpretation of the parameter $\sigma^2$ in the minimax weights \eqref{eq_balancing-weights} is that it controls a bias-variance trade-off. A small $\sigma^2$ emphasizes balance and therefore control of bias and a large $\sigma^2$ emphasizes control of the magnitude of the weights and therefore of variance. This parameter is often selected on the grounds of substantive knowledge or intuition.
For example, in the context of a medical study, a domain expert may tell us that the treatments groups are comparable in terms of age if they don't differ by more than a few months.\footnote{ It can be more intuitive to minimize the magnitude of the weights subject to a constraint on imbalance, which is equivalent. There exists some data-dependent choice $\sigma^2$ that yields a solution of the constrained problem for each  bound on the imbalance and vice versa \citep[Chapter 5]{boyd2004convex}.}

Varying $\sigma^2$ does control the bias-variance trade-off,
but it does so to a limited extent if precautions are taken to avoid bias.
For rich models,
balancing weights (e.g., \ref{eq_balancing-weights-comparison}) 
converge to the inverse propensity weights, 
which is necessary to control bias and determines variance. 

Extreme choices of $\sigma^2$ can nonetheless have considerable impact, especially in small samples. If we choose $\sigma^2$ to be much larger than the imbalance could ever be, then our weights will all be essentially zero. While there is a range of choices that will result in similar and typically good behavior, it is helpful to have some heuristic to help us make that decision. What we suggest is one arising from the minimax interpretation of the weights discussed in Section~\ref{sec:bal_weights_intro}: 
that we choose $\sigma^2$ to be roughly the conditional variance of the outcomes, $\var[Y_i \mid X_i, W_i]$. This choice is supported by asymptotic theory \citep[e.g.,][]{hirshberg2018augmented, hirshberg2019minimax}, which suggests that taking $\sigma^2$ to be constant order (not scaled as a function of sample size), is a relatively robust choice. 

How to select this parameter automatically is an open question. Some approaches have been proposed for selecting it automatically \citep{kallus2020generalized, wang2020minimal,zhao2019covariate} but theoretical and empirical evidence supporting them is currently limited.

%

\section{Other estimands}
\label{sec:other-estimands}

Thus far, we have considered estimating the average treatment effect $\tau = \mu_1 - \mu_0$, focusing on estimating $\mu_1$ in particular.
These ideas and this framework, however, can be used for many other quantities of interest.
In particular, they generalize
to estimate
many other functions $\estimand(m)$ of the conditional mean outcome $m_w(x) = \E[Y \mid X=x,\ W=w]$, where the treatment $w$ need not be discrete-valued. This includes many generalizations of the average treatment effect. For example,
\begin{align} 
&\estimand(m) = \E[ m_{\pi(X)}(X) ], &&\text{average outcome with dosing policy $w=\pi(x)$}; \\
&\estimand(m) = \E\sqb*{\frac{\partial}{\partial w} m_{w} (X) \mid_{w=W} }, &&\text{average effect of an infinitesimal dose increase}; \\
&\estimand(m) = \int \{m_1(x) - m_0(x)\} p(x)dx, &&\text{population-targeted average treatment effect}. \label{eq:targeted-ate}
\end{align}
All of these are instances of a  general estimand that can be estimated with essentially the same approach
\citep[see, e.g,][]{chernozhukov2018biased, hirshberg2018augmented}.
\begin{equation}
\label{eq:amle-general}
\estimand(m) = \E[ h(W_i, X_i, m) ]  \quad \text{ where } \quad h(x,w,m) \ \text{ is linear in }\ m.
\end{equation}
Above, where we have focused on the case that $\estimand(m) = \mu_1 = \E[m_1(X_i)]$, the inverse propensity weights $\gamma_\estimand(w,x)=w\gammaipw(x)$ have played an essential role. In general, this role is played by the \emph{Riesz representer} of the function $\estimand$, which is the solution $\gamma_{\estimand}(w, x)$ to a balance condition that generalizes the one defining the inverse propensity weights \eqref{eq_mipw-balance}:
\begin{equation}
\label{eq:riesz-representer-general}
\E[h(W_i,X_i, f)] = \E[ \gamma_{\estimand}(W_i, X_i)f(W_i,X_i)] \ \text{ for all }\  f \ \text{ with }\ \E[ f(W_i, X_i)^2 ] < \infty. 
\end{equation}
The solution $\gamma_\estimand$ is guaranteed to exist and be square integrable when $\estimand$ is \emph{mean-square-continuous}. For $\estimand(m)=\mu_1$, mean-square continuity is 
 equivalent to sufficient overlap.

We can mechanically derive generalized forms of everything we discuss above: the IPW and AIPW estimators, imbalance-focused decompositions of their error, appropriate forms of maximal imbalance, and estimators for balancing weights.
To do this, we substitute
$\gamma_{\estimand}(W_i,X_i)$ for $W_i\gammaipw(X_i)$ and then
$h(W_i,X_i,m)$ for $m_1(X_i)$ in Equations~\ref{eq_general_ipw}-\ref{eq_balancing-weights-general} above.%
%
\footnote{Order is important when making these substitutions only because we have followed convention in writing $W_i m_1(X_i)$ where the equivalent expression $W_i m_{W_i}(X_i)$ is perhaps more natural.} 
All of our discussion above generalizes, whether focused more on maximal imbalance 
as in Sections~\ref{sec:balancing-binary}-\ref{sec:role-of-augmentation} 
\citep{hirshberg2018augmented} or on estimation of the inverse propensity score as in Section~\ref{sec:beyond-the-plug-in}  \citep{chernozhukov2018biased}.

Furthermore, a small tweak allows us to estimate some functions $\estimand(m)$ of the conditional mean outcome that are not mean-square continuous, like the conditional average treatment effect (CATE) $\tau(x_0) = m_1(x_0)-m_0(x_0)$ at a point $x_0$. The CATE is an instance of the population-targeted average treatment effect \eqref{eq:targeted-ate} in which the target population's covariate distribution $p(x)$ is a spike $\delta_{x_0}(x)$ at $x=x_0$.
Here, the lack of overlap of this spiky target distribution $\delta_{x_0}$ with the distribution of the observed covariates causes the function not to be continuous.
However, we can approximate the spike with a smoother density, 
taking $p(x)$ to be the density of a normal distribution with mean $x_0$ and covariance $h^2 I$ for small \emph{bandwidth} $h$, which will yield good approximation of $\tau(x_0)$ if $\tau$ is continuous at $x_0$. A refined version of this approach can be used for other estimands like the dose-response curve for a continuous-valued treatment, $\mu(w_0)=\E[m_{w_0}(X_i)]=\E[\int m_{w}(X_i) \delta_{w_0}(w)dw]$, 
which fix some aspects of the treatment or population and average over others \citep[e.g.,][]{chernozhukov2018global, kennedy2017non}.

Other parameters identified by moment conditions can be estimated by a similar approach as well, including quantile treatment effects or average treatment effects identified by instrumental variables.
In essence, if a parameter $\theta_0$ satisfies the moment condition $\E[h(W_i, X_i, m, \theta)] = 0$ 
for any function $m$ that we can estimate, and we can propose a valid model $\model$ for this estimator's error  $\dm = \hat m - m$, then for any given $\theta$ we can form a generalized AIPW estimator $\hat\estimand_{\theta}$ of $\estimand_{\theta}(m) = \E[h(W_i,X_i,m,\theta)]$. To estimate $\theta_0$, all we must then do is solve $\hat \estimand_{\theta} \approx 0$. See \citet{chernozhukov2016locally, singh2019biased}.

\ifjasa\else
\section{The connection between balancing and modeling}
\label{sec:dual}

We can characterize balancing weights as regularized M-estimators of the inverse propensity weights using Lagrange duality. This puts focus on the choice of the dispersion measure $\dispersion$, which determines the parameterization of the weights and the loss function they minimize, and the imbalance measure $\imbalancescale$,
which determines the form of the regularization.
In particular, 
for a model $\model$
that is a convex set and various choices of $\dispersion$ and $\imbalancescale$, 
the weights that solve Equation~\ref{eq_balancing-weights-general} 
are equivalently characterized as $\hgamma(X_i) = \eta_{\dispersion}\{g(X_i)\}$ where $g$ solves 
\begin{equation}
    \label{eq:dual_functions}
    \min_{g} \underbrace{\frac{1}{n}\sum_{i=1}^n W_i D_\dispersion\left(\gammaipw(X_i) \ \middle\Vert \  \eta_{\dispersion}\{g(X_i)\}\right)}_{\text{average Bregman divergence}} - \underbrace{\p*{\frac{1}{n}\sum_{i=1}^n g(X_i) - \frac{1}{n}\sum_{i=1}^n W_i \gammaipw(X_i) g(X_i)}}_{\text{mean zero ``noise''}} + \  \vphantom{\sum_{W_i=0}}\imbalancescale^\ast\left(\|g\|_\calM\right).
  \end{equation}
  Here, $\imbalancescale^*(y) = \sup_x x y - \imbalancescale(x)$ is the \emph{convex conjugate} of the function $\imbalancescale$ \citep[see, e.g.,][]{peypouquet2015convex}.
We give a more precise statement and a proof in Appendix~\ref{sec:formal-duality}.
  
From this characterization, we can interpret the weights as a kind of generalized linear model for the inverse propensity score. The function $g$ is the `natural parameter', which is mapped into an estimate of the inverse propensity score by a \emph{link function} $\eta_{\dispersion}(x)=\dispersion^{\ast \prime}(x)$ that is the derivative of the convex conjugate of the dispersion measure $\dispersion$. This model is fit by minimizing a noisy, penalized, asymmetric distance
between the estimate $\hgamma$ and the inverse propensity weights $\gammaipw$ --- specifically, the average \emph{Bregman divergence} associated with $\dispersion$.\footnote{The Bregman divergence is the difference $D_{\dispersion}(x \Vert y)=\dispersion(x) - \dispersion(y) - \dispersion'(y)(x-y)$ between a finite difference and a first order approximation to it. For the square $\dispersion(x)=x^2$, it is squared error $(x-y)^2$.} 
The penalty $\imbalancescale^\ast\left(\|g\|_\model\right)$ acts on the natural parameter $g$, increasing with the 
gauge $\norm{g}_{\model}$.

While different choices of $\dispersion$ and $\imbalancescale$ favor certain interpretations of the weights, we think they play a secondary role from an inferential standpoint.
As discussed in Section~\ref{sec:true_ipw}, when weights balance a rich enough model to adjust for all plausible forms of confounding, they must converge to the inverse propensity weights, so their specific parameterization is a second order concern. We can see this in terms of the dual \eqref{eq:dual_functions} as well. Regardless of the specific form of Bregman divergence $D_\dispersion$, it will induce $\eta_\dispersion\{g(\cdot)\}$ to converge to $\gammaipw$ as long as regularization allows, i.e.,  as long as we can approximate $\gammaipw$ arbitrarily well by $\eta_\dispersion\{g(\cdot)\}$ while  $\imbalancescale^\star(\norm{g}_\model) < \infty$. This is the case when $\eta_\dispersion$ is invertible, $\model$ is nonparametric,
and $\imbalancescale^\star$ is finite except at infinity. 

\subsection{Examples}
We now show some examples of the correspondence between primal and dual weighting problems, discussing the choices of 
$\dispersion$ and $\imbalancescale$ in the primal and the corresponding link function $\eta_{\dispersion}$, Bregman divergence $D_{\dispersion}$, and regularization curve $\imbalancescale^{\star}$ in the dual.
We summarize the correspondences used in our two main examples in the rows of the Table \ref{tab:duality} below. See \citet{Ben-Michael2019}, \citet{wang2020minimal}, and \citet{zhao2019covariate} for related discussions. 

\begin{table}[h!]
    \centering

    \begin{tabular}{cccccc}
$\dispersion(\gamma)$ & $\eta_\dispersion(g)=\dispersion^{\ast \prime}(g)$ & $D_{\dispersion}(x \Vert y)$
& $\imbalancescale(x)$   & $\imbalancescale^\star(y)$ \\
\hline 
$\gamma^2/2$   & $g$ & $(x-y)^2/2$ & $\mathcal{I}( x \le 1)$ & $x$ \\
$\gamma \{\log(\gamma)-1\}$ & $\exp(g)$ & $x \log(x/y) - x + y$
& $\mathcal{I}( x \le 1)$ & $x$ \\
\hline 
    \end{tabular}
    \caption{Correspondence between primal and dual weighting problems}    
    \label{tab:duality}
\end{table}

\subsubsection{The Lasso minimum distance estimator of the inverse propensity score and stable balancing weights}
\label{sec:lasso_min_dist}
As discussed in Section~\ref{sec:bias-variance-tradeoff}, we can equivalently reparameterize the minimax weights to constrain rather than penalize imbalance. For example, we could use the constrained variant of the minimax weights for the linear model $\mathcal{M} = \{m(x) = \beta \cdot \phi(x) : \sum_j \abs{\beta_j}/\lambda_j \le 1\}$,
\begin{equation}
\label{eq:lasso-md-primal}
\begin{aligned}
    \argmin_\gamma \; & \frac{1}{2n}\sum_{i=1}^n W_i \gamma(X_i)^2 
    \quad \text{subject to the constraints} \\ 
    &\abs*{\frac{1}{n}\sum_{i=1}^n W_i\gamma(X_i) \phi_j(X_i) - \frac{1}{n}\sum_{i=1}^n \phi_j(X_i) } \le 1/\lambda_j \ \text{ for all }\ j. \\
\end{aligned}
\end{equation}
Like the minimax weights  \eqref{eq_balancing-weights}, this uses a squared dispersion penalty $\dispersion(\gamma) = \gamma^2/2$.
However, here the imbalance measure $\imbalancescale$ is an indicator function $\mathcal{I}(x \le 1)$ that is zero if $x \le 1$ and
infinity otherwise. 
The dual characterizes these weights as an $\ell_1$-penalized linear least squares estimator of the inverse propensity weights, satisfying $\hgamma(X_i) = \hat\theta \cdot \phi(X_i)$ where $\hat\theta$ solves

\begin{equation}
\label{eq:lasso-md-dual}
\begin{aligned}
    &\argmin_{\theta} \frac{1}{2n}\sum_{i=1}^n W_i \set*{ \theta \cdot \phi(X_i)}^2 - \frac{1}{n}\sum_{i=1}^n \theta \cdot \phi(X_i) + \sum_j \abs{\theta_j}/\lambda_j  \\
    =& \argmin_{\theta} \frac{1}{2n}\sum_{i=1}^n W_i \set*{ \gammaipw(X_i) - \theta \cdot \phi(X_i) }^2 \\
-& \set*{ \frac{1}{n}\sum_{i=1}^n \theta \cdot \phi(X_i) -
        \frac{1}{n}\sum_{i=1}^n W_i \gammaipw(X_i) \theta \cdot \phi(X_i) } + \sum_{j} \abs{\theta_j} / \lambda_j.
\end{aligned}
\end{equation}
The second expression minimized, which differs from the first only by terms that do not vary with $\theta$ and therefore do not affect the argmin, is written in the form of our general dual problem \eqref{eq:dual_functions}.

Taking the dual perspective, \citet{chernozhukov2018biased} proposed these weights as an $\ell_1$-penalized least squares estimator for the inverse propensity score. 
Their interpretation focuses on accuracy in estimating the inverse propensity score --- or its projection onto the basis $\phi(x)$ --- as described in Section~\ref{sec:beyond-the-plug-in}.
They advocated cross-fitting the weights.\footnote{Cross-fitting refers to using in each term of the AIPW estimator \eqref{eq:aipw2} a weight $\hat\gamma^{(-i)}(X_i)$ based on an estimate $\hat\gamma^{(-i)}$ from a subsample that does not include unit $i$.} 
Whether cross-fitting the weights results in an estimator that behaves substantially differently from an analogous estimator without cross-fit weights is an area of open exploration, but the properties ascribed to these sibling estimators based on extant analyses are fairly different. Cross-fitting facilitates analysis along the lines discussed in Section~\ref{sec:beyond-the-plug-in},
which leads to claims about double-robustness,
but hinders the analytic approach described in Section~\ref{sec:plugin}, which leads to claims of a form of  optimality for the variant without cross-fitting. Understanding the actual differences between these estimators is, to our knowledge, an open problem. 

The stable balancing weights \citep{zubizarreta2015stable}
are closely related. In the primal perspective, these weights arise from solving a variant of the primal \eqref{eq:lasso-md-primal} with the additional constraints that the weights are non-negative and average to one: they are the weights of minimum variance that control the maximal imbalance over the model $\model$ 
without extrapolating. 
As discussed in Section~\ref{sec:translation-invariance}, the constraint that the weights average to one can be encoded by use of a translation invariant model, i.e., by including a constant basis function $\phi_j(x)=1$ with slack $1/\lambda_j=0$.
To add the non-negativity constraint we replace the dispersion measure $\dispersion(\gamma)=\gamma^2/2$ with a variant $\dispersion(\gamma)= \mathcal{I}(\gamma \ge 0) \cdot \gamma^2/2$ that is infinite for negative weights. The corresponding dual characterization of the weights is that they follow a generalized linear model
$\hgamma(X_i) = \eta_{\dispersion}(\theta \cdot X_i)$
with the positive-part link $\eta_{\dispersion}(g) =  (g)_+$ and parameter $\theta$ solving 
\begin{equation}
\label{eq:sbw-dual}
\begin{aligned}
 &\argmin_{\theta} \frac{1}{2n}\sum_{i=1}^n W_i \set*{\theta \cdot \phi(X_i)}_+^2 - \frac{1}{n}\sum_{i=1}^n \theta \cdot \phi(X_i) + \sum_j \abs{\theta_j}/\lambda_j  \\
=& \argmin_{\theta} \frac{1}{2n}\sum_{i=1}^n W_i \sqb*{ \gammaipw(X_i) - \set*{\theta \cdot \phi(X_i)}_+ }^2 - \frac{1}{n} \sum_{i=1}^n \set{\theta \cdot \phi(X_i)}_- \\
-& \sqb*{ \frac{1}{n}\sum_{i=1}^n \set*{ \theta \cdot \phi(X_i) }_+ -
        \frac{1}{n}\sum_{i=1}^n W_i \gammaipw(X_i) \set*{\theta  \cdot \phi(X_i)}_+ } + 
        \sum_{j} \abs{\theta_j} / \lambda_j.
\end{aligned}
\end{equation}
Here $x_+ = \max(x,0)$ and $x_-=\min(x, 0)$. 
The first characterization is implied by the more general form of the dual stated in Proposition~\ref{prop:duality} in the appendices. The second characterization, however, does not and can not have the general form of the dual described above \eqref{eq:dual_functions}: Because this choice of $\dispersion$ is not differentiable, the Bregman divergence $D_{\dispersion}$ is not defined in the conventional sense. This characterization does, as in cases that admit a Bregman divergence, include a distance between $\gammaipw$ and a generalized linear model $\eta_{\dispersion}\{g(\cdot)\}$, a mean zero noise term, and a regularization term. It differs in that it includes an additional non-negative term, \smash{$-n^{-1}\sum_{i=1}^n \{\theta \cdot \phi(X_i)\}_-$}, which prevents the generalized linear model's \emph{index} $g(X_i)=\theta \cdot \phi(X_i)$ from becoming too negative.

\subsubsection{Entropy balancing and the regularized calibrated propensity score}

Entropy balancing, proposed by \citet{hainmueller2012balancing}, uses weights solving an instance of the following optimization problem.
\begin{equation*}
\begin{aligned}
    \argmin_\gamma \;  &\frac{1}{n}\sum_{i=1}^n W_i \gamma(X_i) \{\log \gamma(X_i) - 1\} \quad \text{subject to the constraints} \\ &\ \frac{1}{n}\sum_{i=1}^n \abs*{W_i\gamma(X_i) \phi_j(X_i) - \frac{1}{n}\sum_{i=1}^n \phi_j(X_i) } \le 1/\lambda_j \ \text{ for all }\ j. \\
\end{aligned}
\end{equation*}
This is an instance of the primal \eqref{eq_balancing-weights-general} with model $\mathcal{M} = \{m(x) = \beta \cdot \phi(x) : \sum_j \abs{\beta_j}/\lambda_j \le 1\}$, \\
imbalance measure $\imbalancescale(x) = \mathcal{I}(x \le 1)$, and dispersion measure $\dispersion(\gamma) = \gamma(\log \gamma - 1)$.
The name refers to the way these weights are optimized for dispersion: 
what is minimized is the entropy of the weights minus their average (which is a constant if we use a translation invariant model).

In their dual form, these weights have 
been proposed independently as an estimator of the propensity score using a log-linear model that is \emph{calibrated} in the sense that it achieves balance even if misspecified \citep{tan2020regularized}. In particular, the dual characterizes the estimated inverse propensity weights as $\hgamma(X_i) =\exp\{\hat\theta \cdot \phi(X_i)\}$ for $\hat \theta$ solving
\begin{equation*}
\begin{aligned}
&\argmin_{\theta} \frac{1}{n}\sum_{i=1}^n W_i\exp\left\{\theta \cdot \phi(X_i)\right\} - \frac{1}{n} \sum_{i=1}^n \theta \cdot \phi(X_i) + \sum_{j} \abs{\theta_j} / \lambda_j \\
=& \argmin_{\theta} \frac{1}{n}\sum_{i=1}^n W_i \p*{ \gammaipw(X_i) \log\sqb*{ \frac{\gammaipw(X_i)}{\exp\{\theta \cdot \phi(X_i) \}}} -
\gammaipw(X_i) + \exp\{\theta \cdot \phi(X_i)\}} \\
&- \p*{ \frac{1}{n}\sum_{i=1}^n \theta \cdot \phi(X_i) -
        \frac{1}{n}\sum_{i=1}^n W_i \gammaipw(X_i) \theta \cdot \phi(X_i) } + \sum_{j} \abs{\theta_j} / \lambda_j
\end{aligned}
\end{equation*}
The second expression minimized, which differs from the first only by terms that do not vary with $\theta$ and therefore do not affect the argmin, is written in the form of our general dual problem \eqref{eq:dual_functions}. The Bregman divergence $D_\dispersion$ involved is the generalized KL divergence.


This approach tends to be discussed more in the context
of estimation of the ATT, in which case the dual is essentially similar.\footnote{See the more general form of the dual in Proposition~\ref{prop:duality}.}
Because the generalization of the inverse propensity weights $\gammaipw(x)$ for estimating the ATT are proportional to the odds $\{1-e(x)\}/e(x)$, the parameterization $\gammaipw(x)=\exp(\theta \cdot x)$ has an interpretation as a logistic regression model for the propensity score, $e_{\theta}(x) = 1/[1+\exp\{\theta \cdot \phi(x)\}]$, instead of a log-linear one.

\citet{tan2017regularized} showed that with the use of a specialized linear regression estimator $\hat m_1$, the augmented estimator using these weights has an interesting high-dimensional double robustness property. It is asymptotically normal with mean zero and estimable variance if \emph{either} the logistic propensity model $e(x)=1/[1+\exp\{\theta \cdot \phi(x)\}]$ or the linear outcome model $m_1(x) = \beta \cdot \phi(x)$ is correctly specified, so long as there is enough sparsity in the `pseudo-true' parameters $\theta^{\star}$ and $\beta^{\star}$ that the estimates $\hat\theta$ and $\hat \beta$ converge to. This is a high-dimensional generalization of the concept of double robustness that arises when working with classical parametric models
--- that it is acceptable to misspecify one model but not both --- 
that is not captured faithfully by the usual nonparametric 
`product of rates'
concept discussed in Section~\ref{sec:beyond-the-plug-in}. Similar estimators and their double robustness properties
have received considerable attention lately \citep[e.g.,][]{ benkeser2017doubly,bradic2019minimax,bradic2019sparsity,ning2020robust}.

\fi

\singlespacing
\printbibliography

\ifappendix
\clearpage
\appendix
\begin{refsection}

\section{Asymptotics}
\label{sec:appendix-asymptotic-normality}


\begin{proposition}

 \label{proposition:appendix-clt}
Suppose that the noise is well behaved in the sense that its conditional variance $\nu(X_i) := \E[\varepsilon_i^2 \mid X_i, W_i=1]$ is bounded away from zero and that it satisfies a uniform integrability condition
$\max_{i \le n} \E[\varepsilon_i^2 1(\varepsilon_i^2 \ge M) \mid W_1,X_1 \ldots W_n, X_n] \to_p 0$ as $M \to \infty$. If imbalance in $\dm$ is negligible in the sense that it is $o_p\{n^{-1/2}\norm{w\hgamma}_{L_2(\Pn)}\}$ and the weights are infinitesimal in the sense that $\max_{i \le n} W_i \hgamma(X_i)^2 = o_p\{\sum_{i\le n} W_i \hgamma(X_i)^2\}$, then
 \[  \frac{\sqrt{n}(\hat\mu^{\aug} -  \tilde \mu_1)}{\sqrt{\hat V}} \to N(0,1) \quad \text{ where } 
 \quad \hat V = \frac{1}{n}\sum_{i=1}^n W_i \hgamma(X_i)^2 (Y_i - \hat m_1^{(-i)}(X_i))^2 \] 
if $\hat m_1^{(-i)}$ is a $K$-fold cross-fit estimator of $m_1$ that is  consistent in the sense that $\max_{i \le n} \abs{\hat m_1^{(-i)}(X_i) - m_1(X_i)} = o_p(1).$ 
\end{proposition}

 \noindent Here $\tilde \mu_1 = n^{-1}\sum_{i=1}m_1(X_i)$, and where we say $\hat m_1^{(-i)}$ is a $K$-fold cross-fit estimator of $m_1$, we mean that for some partition $I_1 \ldots I_K$ of the indices $1 \ldots n$, $\hat m_1^{(-i)}$ is independent of the observations $\set{(W_{i'}, X_{i'}, Y_{i'}) : i' \in I_k}$ where $I_k$ is the element of the partition containing $i$.


\begin{proof}
This proof has two steps. We'll show that  $Z_n=\sqrt{n}(\hat\mu^{aug} -  \tilde\mu)/V_n^{1/2}$ is asymptotically standard normal for $V_n = n^{-1}\sum_{i=1}^n W_i \hgamma(X_i)^2 \nu(X_i)$, then we'll show that $\hat V / V_n$ converges to one in probability, so Slutsky's theorem implies our claim. For the first step, we work with the decomposition
\[ 
\sqrt{n}(\hat \mu_1^{aug} - \tilde\mu_1) = 
\sqb*{ \frac{1}{\sqrt{n}}\sum_{i=1}^n W_i\hat{\gamma}(X_i) \dm(X_i) - \frac{1}{\sqrt{n}}\sum_{i=1}^n \dm(X_i) } + 
\sqb*{\frac{1}{\sqrt{n}}\sum_{i=1}^n  W_i\hat{\gamma}(X_i)\varepsilon_i  }.
\]
The first term is $o_p(\norm{ w \hgamma}_{L_2(\Pn)})$ by assumption and therefore, because $\nu$ is bounded away from zero, negligible relative to the denominator $V_n^{1/2}$, which excedes $\norm{w \hgamma}_{L_2(\Pn)} \underline{\nu}^{1/2}$ for $\underline{\nu}=\min_{x}\nu(x)$. 
It follows that $\sqrt{n}(\hat \mu_1^{aug} - \tilde\mu_1)/V_n^{1/2}$ converges in probability to  $V_n^{-1/2}$ times the second term.
Conditional on the ning $\mathcal{D}_n=\set{W_i, X_i : i \le n}$, this is a standardized sum of independent mean zero random variables
$\xi_{i,n}= W_i\hat{\gamma}(X_i)\varepsilon_i / (n V_n)^{1/2}$, so 
the central limit theorem implies it converges to a standard normal random variable if the sum is not dominated by too few terms.
To be more precise, letting $\hat\gamma$
refer to the weights estimated at sample size $n$ when defining $\xi_{i,n}$, this is a martingale difference array adapted to the $\sigma$-algebras $\mathcal{F}_{i,n}$ generated by $\mathcal{D}_n \cup \set{\varepsilon_1 \ldots \varepsilon_{i}}$, it is standardized in the sense that its sum of conditional variances $\sum_{i=1}^n \E[\xi_{i,n}^2 \mid \mathcal{F}_{i-1,n}]$ is one, and this conditional Lindeberg condition suffices to ensure its asymptotic normality \citep[Corollary 3.1]{hall2014martingale}:
\[ \sum_{i=1}^n \E[\xi_{i,n}^2 1(\xi_{i,n}^2 \ge \epsilon) \mid \mathcal{F}_{i-1,n}] \to_p 0 \quad \text{ for all } \epsilon > 0. \]
To show that our assumptions imply this Lindeberg condition, we decompose it and use H\"older's inequality to bound it as follows.
\begin{align*}
    \sum_{i=1}^n \E[\xi_{i,n}^2 1(\xi_{i,n}^2 \ge \epsilon) \mid \mathcal{D}_n] &=  \frac{1}{nV_n}\sum_{i=1}^n W_i \hgamma(X_i)^2 \E[\varepsilon_i^2 1(\varepsilon_i^2 \ge M) \mid \mathcal{D}_n] \\& +
    \frac{1}{nV_n}\sum_{i=1}^n W_i \hgamma(X_i)^2 \E[\varepsilon_i^2 1(M \ge \varepsilon_i^2 \ge nV_n) \mid \mathcal{D}_n] \\
&\le \frac{1}{n V_n}\sum_{i=1}^n W_i \hgamma(X_i)^2 \cdot  \max_{i \le n}\E[\varepsilon_i^2 1(\varepsilon_i^2 \ge M) \mid \mathcal{D}_n]\\
&+ \frac{1}{nV_n}\sum_{i=1}^n W_i \hgamma(X_i)^2 \cdot  M\max_{i \le n}1(W_i \hgamma(X_i)^2 M \ge nV_n).
\end{align*}
Both terms in this bound have the same leading factor, the ratio of $n^{-1}\sum_{i=1}^n W_i \hgamma(X_i)^2$ and $n^{-1}\sum_{i=1}^n W_i \hgamma(X_i)^2 \nu(X_i)$, which is $O_p(1)$ because $\nu$ is bounded away from zero. Thus, it suffices to show that the sum of the second factors is $o_p(1)$. Our uniform integrability assumption implies that for large enough $M$, the probability that the first term's exceeds any threshold $\delta>0$ is arbitrarily small.  
Moreover, for this $M$, the second term's tends to zero with probability tending to one, as given our assumption that $\nu$ is bounded away from zero, the claim that
$\max_{i \le n}W_i\hgamma(X_i)^2 / \sum_{i \le n}W_i \hgamma(X_i)^2 \nu(X_i) \to_p 0$
reduces to our assumption that the weights are infinitesimal. This concludes the first step of our proof.

For the second step, it is equivalent to show that $(\hat V - V_n)/V_n \to 0$ in probability.
To do this, we work with the decomposition
\begin{align*} 
\frac{\hat V - V_n}{V_n}
&= \frac{1}{n}\sum_{i=1}^n \alpha_i [ \{\varepsilon_i + m_1(X_i) - \hat m_1^{(-i)}(X_i)\}^2 - \nu(X_i)] \ \text{ for } \ \alpha_i = \frac{W_i \hgamma(X_i)^2}{V_n} \\
&= \frac{1}{n} \sum_{i=1}^n \alpha_i \{m_1(X_i) - \hat m_1^{(-i)}(X_i)\}^2 + \frac{2}{n} \sum_{i=1}^{n} \alpha_i \varepsilon_i \{m_1(X_i) - \hat m_1^{(-i)}(X_i)\} \\
&+
\frac{1}{n} \sum_{i=1}^n \alpha_i \{\varepsilon^2 - \nu(X_i)\}. 
\end{align*}
We will show that each term converges to zero. Note that $n^{-1}\sum_{i=1}^n \alpha_i = \norm{w\hgamma}_{L_2(\Pn)}^2/V_n$ is, as discussed above, bounded because $\nu$ is bounded away from zero.
\begin{enumerate}
\item
By H\"older's inequality, 
the magnitude of the first term is bounded by $n^{-1}\sum_{i=1}^n \alpha_i \cdot \max_{i \le n}\{m_1(X_i)-\hat m_1^{(-i)}(X_i)\}^2$.  Because the first factor is bounded, this converges to zero in probability if the second does,
and therefore under our consistency assumption.

\item Let $I_1 \ldots I_K$ be the partition of the indices $i=1 \ldots n$
used to cross-fit $\hat m_1^{(-i)}$,
so for $i \in I_k$, $\hat m_1^{(-i)}$
is independent of the observations for $i \in I_k$. We can write the second term as the double sum $(2/n)\sum_{k=1}^K \sum_{i \in I_k}\alpha_i \varepsilon_i \{ m_1(X_i) - \hat m_1^{(-i)}(X_i)\}.$ Conditioning on
everything but the outcomes for $i \in I_k$,
i.e., on the $\sigma$-algebra generated by $\set{(W_i,X_i,Y_i) : i \in I_k^c} \cup \set{ (W_i,X_i) : i \in I_k}$, Chebyshev's inequality
implies that each inner sum satisfies the following bound with conditional probability $1-\delta$:
\[ \abs*{\sum_{i \in I_k}\alpha_i \varepsilon_i \{ m_1(X_i) - \hat m_1^{(-i)}(X_i)\}} \le   \sqrt{\delta^{-1}\sum_{i \in I_k} \alpha_i^2 \nu(X_i)\{m_1(X_i) - \hat m_1^{(-i)}(X_i)\}^2}. \]
It follows that each holds with unconditional probability $1-\delta$ and, via the union bound, that all of these hold simultaneously on an event of probability $1-K\delta$. On this event, the sum itself is bounded by $K$ times the largest one of these, and therefore
\begin{align*} \abs*{(2/n)\sum_{i=1}^n\alpha_i \varepsilon_i \{ m_1(X_i) - \hat m_1^{(-i)}(X_i)\}} 
&\le  \sqrt{\frac{4K^2}{\delta n^2}\sum_{i=1}^n \alpha_i^2 \nu(X_i)\{m_1(X_i) - \hat m_1^{(-i)}(X_i)\}^2} \\
&\le \sqrt{\frac{4K^2}{\delta n}\max_{i \le n} \alpha_i \nu(X_i) \cdot \frac{1}{n}\sum_{i \le n} \alpha_i \{m_1(X_i) - \hat m_1^{(-i)}(X_i)\}^2}.
\end{align*}
The second bound follows from the first via H\"older's inequality. In it, the first factor is bounded by $4K^2/\delta$ times $n^{-1}\sum_{i=1}^n \alpha_i \nu(X_i) = V_n/V_n=1$ and the second factor is the first term in our decomposition, which we have shown converges to zero above. 
\item The third term is difference between the sum of squares $\sum_{i=1}^n \xi_{i,n}^2$ and the sum of conditional variances $\sum_{i=1}^n \E[\xi_{i,n}^2 \mid \mathcal{F}_{i-1,n}]$ of the standardized martingale difference array $\xi_{i,n}$, and the aforementioned conditional Lindeberg condition is sufficient to imply that the two converge in probability \citep[Theorem 2.23]{hall2014martingale}.
\end{enumerate}
\end{proof}

\section{More on Models}
\label{sec:more-on-models}

\subsection{Sobolev Space Details}
\label{sec:sobolev-spaces}
The particular norm we state in Equation~\ref{eq:sobolev-derivative} is the norm of the order-1 mixed smoothness Sobolev space of bivariate functions on the cube $[0, \pi]^2$ \citep[e.g.,][Section 2.1]{kuhn2015approximation}. Sometimes this is described as a norm on 
functions on $[-\pi,\pi]^2$ that are $2\pi$ periodic in all coordinates, and in Equation~\ref{eq:sobolev-derivative}
the average over the uniform distribution on
$[0,\pi]^2$ is replaced with an average over the uniform distribution over $[-\pi,\pi]^2$; the arbitrary functions on $[0,\pi]^2$ can be extended to such functions on $[-\pi,\pi]^2$ via `reflection', letting $f(\pm x_1, \pm x_2)=f(x_1, x_2)$.
For these `reflections', the norm in Equation~\ref{eq:sobolev-derivative} and the variant that averages over $[-\pi,\pi]^2$ match.

\subsection{Models defined in terms of $\ell_q$ and $L_q$ norms}
\label{sec:lq-models}
In Section~\ref{sec:balance_models}, we have discussed models defined in terms of $\ell_2$ norms on coefficients (\ref{eq:bias_linear}-\ref{eq:imbalance_phi_infinite}). It is common to use models defined analogously in terms of the $\ell_1$ norm, or another $\ell_q$ norm, on the coefficients. For example, the
largest imbalance in any coordinate $X_{ij}$ is the maximal imbalance for a variant of the model \eqref{eq:bias_linear} using the $\ell_1$ norm and
the sum of absolute imbalances in the coordinates arises from a model based on the $\ell_{\infty}$ norm.
In models with a low-dimensional basis, such as models without interactions, these are roughly equivalent, in the sense that the unit balls of the $\ell_1$, $\ell_2$, and $\ell_{\infty}$ norms on $\R^p$ satisfy $p^{-1} B_{\infty} \subseteq p^{-1/2} B_2 \subseteq B_1 \subseteq B_2 \subseteq B_{\infty}$.
For models requiring a high dimensional basis ($p$ large), the same relationship does not spell equivalence. Geometrically, $B_1$ is a `diamond' inscribed in $B_2$, a `round ball', inscribed in $B_{\infty}$,
the `cube' with side-length $2$. In $\R^2$ these are of increasing but similar size, whereas in high dimensional spaces almost all the volume of $B_{\infty}$ is in the corners, outside $B_2$, and similarly almost all the volume of $B_2$ is outside of $B_1$. In fact, the size of $B_1$ is, in a meaningful sense, almost independent of dimension $p$. This fact lies at the heart of the tractability of some problems in high dimensional statistics and their connection to $\ell_1$ regularization, including the problem of balancing high dimensional linear models \citep[e.g.,][]{athey2018approximate, tan2020regularized}.

The RKHS model discussed above \eqref{eq:imbalance_phi_infinite} is based on a \emph{weighted} $\ell_2$ norm on infinitely many scaled coefficients $\beta_j / \lambda_j$.
When the scale factors $1/\lambda_j$ increase sufficiently fast, the corresponding coefficients $\beta_j$ must decay rapidly, and as a result this model behaves like one defined in terms of a low-dimensional basis. In particular, models defined analogously in terms of other $\ell_q$ norms on $\beta_j / \lambda_j$ are of similar size.

Similarly, generalizing the Sobolev balls we have discussed above, based on $L_2$ norms on derivatives \eqref{eq:sobolev-derivative}, are Sobolev balls defined analogously in terms of other $L_q$ norms. These are of comparable size for different $q$ if the functions in them are sufficiently smooth, e.g., $L_2$ and $L_\infty$ Sobolev balls of functions on $\R^p$ are of similar size if we include derivatives of order greater than $p/2$ in the norm \citep[e.g.,][Corollary 4]{nickl2007bracketing}. 

\subsection{Models defined in terms of variation norms}
\label{sec:variation-models}
To check balance in covariate distributions, researchers will sometimes compare histograms of the covariates for the re-weighted treated units to those of the population, often summarizing this comparison by the marginal Kolmogorov-Smirnov (KS) statistics.
This measure of imbalance corresponds to an \emph{additive model} for the outcome, $m_1(x)=\sum_{j=1}^d m_{1j}(x_j)$, with a component function $m_{1j}(x_j)$ for the $j$th covariate. In particular, the maximal KS statistic is the maximal imbalance for 
an additive model with a bounded summed componentwise \emph{total variation}, 
$\model = \set{ \sum_{j=1}^d f_j(x_j) : \sum_{j=1}^d \norm{f_j}_{TV} \le 1}$ where, for differentiable functions, $\norm{f_j}_{TV} = \sum_j \int \abs{f_j'(x)}dx$.\footnote{See \cite{fang2021multivariate} for a thorough discussion of total variation. 
One useful interpretation is that $\norm{f}_{TV}$ is the vertical distance a pen travels when drawing the function $f$: for differentiable functions, it is $\int \abs{f'(x)}dx$
and the seminorm associated with the model is $\sum_j \abs{f_j'(x)}dx$.} 
This is a more general approach to controlling bias than comparing covariate means, but it fails to control potential bias arising from imbalance in covariate interactions. This additive model is, however, essentially contained in a very natural model that does include interactions: the unit ball of the space of functions with bounded \emph{sectional variation} \citep{bibaut2019fast, fang2021multivariate}. The norm for this space is similar to the mixed-smoothness Sobolev norm from Equation~\ref{eq:sobolev-derivative}, depending on the same sum of derivatives, but evaluates it via mix of $L_1$-type averaging and evaluation at zero, so it does not coincide with any Sobolev norm exactly \citep[see Equations 23 and 25 in][]{fang2021multivariate}.

\section{Computation in Reproducing Kernel Hilbert Spaces}
\label{sec:RKHS-computation}
Choosing as the model the unit ball $\calB_\calH$ of an RKHS $\calH$ is computationally convenient due to the so-called \emph{kernel trick}.
What lies behind it is the so-called \emph{reproducing property}: for each point $x$, there is an element $K(x,\cdot) \in \calH$ with the property that we can evaluate functions $f \in \calH$ at the point $x$ by taking their inner product with $K(x,\cdot)$, i.e., with the property that $f(x)=\langle K(x,\cdot), f\rangle$ for all $f \in \calH$. As a result, if we take regularized optimization problems over $f \in \calH$ and rewrite each instance of $f(X_i)$ as an inner product with $K(X_i,\cdot)$, it often becomes clear that the solution must lie in the span of $K(X_1,\cdot) \ldots K(X_n, \cdot)$.
Thus, knowing the solution has the form $f(\cdot)=\sum_{i=1}^n \alpha_i K(X_i, \cdot)$ for $\alpha \in \R^n$, we can solve our optimization problem by optimizing over a finite-dimensional vector $\R^n$ instead of an infinite-dimensional function $f \in \calH$.
This is summarized in so-called \emph{representer theorems}. Taking this approach, we can simplify the maximal imbalance over the ball $\calB_\calH$ in terms of its kernel $K$ as follows.

\begin{equation}
\label{eq:imbalance_rkhs}
\imbalance_{\calB_\calH}(\gamma)^2 = \frac{1}{n^2}\sum_{j=1}^n \sum_{i=1}^n K(X_i,X_j) - 2 W_i K(X_i,X_j) \gamma_i +  W_iW_j K(X_i,X_j) \gamma_i\gamma_j.
\end{equation}
\noindent This is helpful when we know the kernel of our space $\calH$. 
See, e.g., \citet{novak2018reproducing} for the kernels associated with
certain Sobolev spaces and techniques for calculating kernels more generally,
and \citet[Section 2.3]{hirshberg2019minimax} and references therein for additional detail on RKHSes more generally.

\section{Augmentation and strong notions of convergence}
\label{sec:strong-norm-consistency}

In Section~\ref{sec:role-of-augmentation}, we discussed the impact of augmentation on
imbalance by focusing on mean-square consistency of $\hat m$ and the equicontinuity of
the imbalance process $f \to n^{-1}\sum_{i=1}^n f(X_i) - W_i \gammaipw(X_i) f(X_i)$.

An argument that appeared earlier  \citep[e.g., in][]{athey2018approximate, kallus2020generalized, wong2018kernel} 
relied on the assumption that $\hat m$ was consistent in a stronger sense: 
instead of requiring the mean squared error $\norm{\dm}_{L_2(P)}$ to vanish, it required the gauge of the error $\norm{\dm}_{\model}$ to do so.
In other words, it required that $\dm$ be in a vanishingly small scaled-down version of $\model$, $\model_n = r_n \model$ for $r_n \to 0$;
in the case of the model $\model = \set{ \beta \cdot x : \norm{\beta}_1 \le 1}$ \eqref{eq:ipw_linf_imbalance}, this means that we must use an estimate of the form
$\hat m_1(x) = \hat\beta \cdot \phi(x)$, and the coefficients must be consistent in the $\ell_1$ norm, as $\norm{\hat m_1-m_1}_{\model} = \norm{\hat\beta-\beta}_1$.
When we have consistency in this sense, $\model_n$ is a valid model for $\dm$,
and the maximal imbalance $\imbalance_{\model_n}(\hgamma) = r_n \imbalance_{\model}(\hgamma)$ is negligible
if $\imbalance_{\model}(\hgamma)$ is $O_p(n^{-1/2})$;
if instead $\imbalance_{\model}(\hgamma)$ is $O_p(\sqrt{\log(p) / n})$ as discussed above for this model,
it suffices to have gauge (e.g., $\ell_1$) consistency with an $o_p(\sqrt{1/\log(p)})$ rate. 
While this argument is perhaps easier to internalize, it is often preferable to use the one above,
as  this stronger form of consistency can distract us with thinking about extraneous aspects of the data. 
For example, multicollinearity of the basis functions $\phi_j$ 
has no effect on our ability to estimate the conditional mean \emph{function} $m_1(x)=\beta \cdot \phi(x)$,
but it can make it impossible to accurately estimate its \emph{coefficients} $\beta$
\citep[see, e.g.,][]{chatterjee2013assumptionless}.

\ifjasa

\fi

\section{Formal Duality Results}
\label{sec:formal-duality}

In this section, we prove a duality result the balancing weights optimization problem \eqref{eq_balancing-weights-general},
in a general form that addresses the general estimand \eqref{eq:amle-general} considered in Section~\ref{sec:other-estimands}.

\begin{proposition}
\label{prop:duality}
Let $\model$ be an absolutely convex set with the property that,
for $i=1 \ldots n$, $\sup_{m \in \model}m(Z_i)$ and $\sup_{m \in \model}h(Z_i,m)$ are bounded for some linear function $h(Z_i,\cdot)$.
Let $\dispersion_i:\R \to \R$ for $i=1\ldots n$ and $\imbalancescale: \R \to \R$ be proper, convex, and lower-semicontinuous functions that are continuous at zero with $\dispersion_i(0)=\imbalancescale(0)=0$, $\dispersion_i$ coercive with $\dispersion_i^\star$ either uniformly convex or both strictly convex and continuous, and $\imbalancescale$ nondecreasing. Define
\begin{align*} 
p(\gamma) &:= \sum_{i=1}^n \dispersion_i(\gamma_i) + \imbalancescale\set*{\sup_{m \in \model} h(Z_i, m) - \gamma_i m(Z_i)}, \\
   d(g) &:= \sum_{i=1}^n \dispersion_i^{\star}\set*{g(Z_i)} - \sum_{i=1}^n h(Z_i, g) + \imbalancescale^{\star}(\norm{g}_{\model}).
\end{align*}
\begin{enumerate}
    \item $p$ has a unique minimum at a vector $\hgamma \in \R^n$.
    \item It satisfies $\hgamma_i \in \partial \dispersion_i^\star\{\lim \hat g_j(Z_i)\}$ for every sequence $\hat g_j$ for which $\lim d(\hat g_j) = \inf_{g}d(g)$.
\end{enumerate}
\end{proposition}
\noindent 
For simplicity, we've used boundedness assumptions that rule out translation invariant 
models as in Section~\ref{sec:translation-invariance}. 
For the treatment specific mean and the problems considered in Section~\ref{sec:other-estimands}, we take $Z_i=(W_i,X_i)$. In the case of the treatment-specific mean, we write the primal with a sum over $i$ with $W_i=1$, not all $i$, so this does not apply as written. A simple generalization of the proof's notation would suffice to address this case; furthermore, essentially as argued in \citet[Section D.3]{hirshberg2019minimax}, because $\gamma_i$ where $W_i = 0$ do not affect balance, so long as $\dispersion_i(x) > 0$ for $x > 0$, and we can assume this without loss of generality because those weights do not affect the estimator.

Equation~\ref{eq:dual_functions} from Section~\ref{sec:dual}  is equivalent when $\dispersion_i=\dispersion$ is differentiable. We will derive this now using a dual characterization of the Bregman divergence:
\[
D_{\dispersion}(a \Vert b) = D_{\dispersion^\star}\{\dispersion'(b), \dispersion'(a)\} =
    \dispersion^{\star}\{ \dispersion'(b) \} - \dispersion^{\star}\{\dispersion'(a)\} - {\dispersion^{\star}}'\{\dispersion'(a)\}\{\dispersion'(b) - \dispersion'(a)\}.
\]
In this expression, take 
$a=\gammaipw(Z_i)$
and $b=\dispersion'^{-1}\{g(Z_i)\}$,
and observe that because ${\dispersion^{\star}}' = {\dispersion'}^{-1}$,
${\dispersion^{\star}}'\{\dispersion'(a)\} = a = \gammaipw(Z_i)$
and $b={\dispersion'}^{-1}(g(Z_i))=\eta_\dispersion(g(Z_i))$ for $\eta(g)={\dispersion^{\star}}'$.
Rearranging, substituting, and collecting terms that don't vary with $g$ in a constant $c$,
\[
\dispersion^{\star}\{g(Z_i)\} 
=  D_{\dispersion}[\gammaipw(Z_i),\  \eta_{\dispersion}\{g(Z_i)\}] 
   +  \gammaipw(Z_i)g(Z_i) + c.
 \]
Substituting this in the first term of $d(g)$ from Proposition~\ref{prop:duality} yields the loss minimized in Equation~\ref{eq:dual_functions}, plus a constant $c$ that does not affect the argmin.

We base our proof of Proposition~\ref{prop:duality} on that of the duality result in \citet[Lemma 5]{hirshberg2018augmented}, which focuses on the special case of the minimax weights.
\begin{proof}

Let $\calS$ be the space of functions with normed by model's gauge $\norm{\cdot} := \norm{\cdot}_{\model}$, 
$H \in \calS^\star$ be the map $H(f) = \sum_{i=1}^n h(Z_i, f)$, and $A:\R^n \to \calS^\star$ be the map
$A\gamma = -\sum_{i=1}^n \gamma_i \delta_{Z_i}$ where $\delta_{Z_i}(f) = f(Z_i)$. Our weights minimize the primal,
\begin{equation} 
p(\gamma) := \sum_{i=1}^n \dispersion_i(\gamma) + \imbalancescale\p*{\norm{H +  A\gamma}_{\star}}. 
\end{equation}
$p$ is proper, convex, coercive, and lower-semicontinuous, so
it has a unique minimum at some $\hgamma \in \R^n$. Its
Fenchel-Rockefellar dual, which for $s(\gamma)=\sum_{i=1}^n \dispersion(\gamma)$ and $r(L)=\imbalancescale(\norm{H + L}_\star)$ is
\begin{equation}
    d(L^\star) := s^{\star}(-A^\star L^\star) + r^\star(L^\star) \quad \text{mapping} \quad \calS^{\star\star} \to \R,
\end{equation}
has a minimum, and each argmin $\hat L^\star$ satisfies the first order optimality condition $-A^\star \hat L^\star \in \partial s(\hgamma)$ where $A^\star$ is the adjoint of $A$ and $s^{\star}$ and $r^\star$ the convex conjugates of $s$ and $r$. Furthermore, $-A^\star \hat L^\star \in \partial s(\hgamma)$ iff $\hgamma \in \partial s^{\star}(-A^\star \hat L^\star)$ \citep[Theorems 2.19 and 3.51 and Proposition 3.59]{peypouquet2015convex}.


We conclude by characterizing $\hat L$ more concretely. Because convex conjugation and summation commute, 
$s^{\star} = \sum_{i=1}^n \dispersion_i^{\star}(\inner{e_i, \cdot})$ for standard basis vectors $e_1 \ldots e_n$ and therefore
$\partial s^{\star} = \sum_{i=1}^n \partial \dispersion_i^{\star}(\inner{e_i, \cdot})e_i$ by the sum and chain rules for subgradients \citep[Theorem 3.30 and Proposition 2.28]{peypouquet2015convex}. It follows that
the first order optimality condition is $\hgamma_i \in \partial \dispersion_i^{\star}(\inner{-A^\star \hat L^\star, e_i}) = \partial \dispersion_i^{\star}(\inner{\hat L^\star, \delta_{Z_i}})$.
Furthermore,    
\begin{align*}
    r^{\star}(L^\star) &= \sup_{L \in \calS^\star} \inner{L^\star, L} - \imbalancescale\p*{\norm{H + L}_{\star}} \\
                       &= - \inner{L^{\star}, H} + \sup_{L' \in \calS^\star} \inner{L^\star, L'} - \imbalancescale\p*{\norm{L'}_{\star}} \\
                       &= - \inner{L^{\star}, H} + \sup_{t \in \R}\sup_{\norm{L''}_{\star}=1} t\inner{L^\star, L''} - \imbalancescale(t) \\
                       &= - \inner{L^{\star}, H} + \sup_{t \in \R}t\norm{L^\star}_{\star\star} - \imbalancescale(t) \\ 
                       &= - \inner{L^{\star}, H} + \imbalancescale^{\star}\p*{\norm{L^\star}_{\star\star}}. \\
\end{align*}            
Here we've used a series of reparameterizations $L=L'-H=tL''-H$ and then definitions of the bidual norm $\norm{\cdot}_{\star\star}$
and the convex conjugate $\imbalancescale^\star$. Substituting these concrete forms,
\begin{equation}
    d(L^{\star}) = \sum_{i=1}^n \dispersion_i^{\star}\p*{\inner{L^\star, \delta_{Z_i}}} - \sum_{i=1}^n \inner{L^{\star}, h(Z_i, \cdot)} + \imbalancescale^{\star}\p*{\norm{L^\star}_{\star\star}}.
\end{equation}
In the particular case that $L^{\star}$ is in the set of evaluation functionals in the bidual, $\set{ E_g(L) = L(g) : g \in \calS} \subseteq \calS^{\star\star}$, this reduces to the claimed dual. While this may be a strict subset, Goldstine's theorem implies that it contains a sequence $E_{\hat g_j}$ along which $\hat g_j(Z_i) \to \inner{\hat L^\star, \delta_{Z_i}}$ and $d$ converges to its minimum $d(\hat L^\star)$ \citep[Theorem 2.6.26]{megginson2012introduction}. We can restate our first order optimality condition as $\hgamma_i \in \partial \dispersion_i^\star\{\lim_j \hat g_j(Z_i)\}$ in terms of this sequence.

We conclude by showing that if $g_j$ is any sequence along which $d$
converges to its minimum, $g_j(Z_i) \to \hat g_j(Z_i)$ and therefore $\hgamma_i \in \partial\dispersion_i^\star\{\lim_j \hat g_j(Z_i)\}$. 
This is because $d$ is in a sense strictly convex. Letting $\tilde g_j$
be the midpoint sequence $(g_j + \hat g_j)/2$, consider  
\begin{equation}
\label{eq:midpoint-convex-d}
\begin{aligned}
d(E_{g_j}) + d(E_{\hat g_j}) - 2d(E_{\tilde g_j}) &=
\sum_{i=1}^n  \sqb*{\dispersion_i^{\star}\{g_j(Z_i)\} + \dispersion_i^{\star}\{\hat g_j(Z_i)\} - 2\dispersion_i^{\star}\{\tilde g_j(Z_i)\}}\\
&+ \sqb*{\imbalancescale^{\star}(\norm{g_j}) + \imbalancescale^{\star}(\norm{\hat g_j}) - 2\imbalancescale^{\star}(\norm{\tilde g_j}) }. 
\end{aligned}
\end{equation}
The term involving $\imbalancescale^{\star}$ is nonnegative because the composition $\imbalancescale^{\star}(\norm{\cdot})$ is convex --- it differs from the convex conjugate $r^{\star}$ by a linear term. Similarly, every
term in the sum is nonnegative because $\dispersion_i^{\star}$ is convex.
Furthermore, if $g_j(Z_i) \not \to \hat g_j(Z_i)$ for some $i$,
the corresponding term in the sum is bounded away from zero for uniformly convex $\dispersion_i^{\star}$. It follows that the whole expression is bounded away from zero, which cannot be the case if $g_j$ and $\hat g_j$ are both minimizing sequences for $d$.

We can adapt this argument to the case that $\dispersion_i^\star$ is strictly convex and continuous. For $g_j$ to be a minimizing sequence, $g_j(Z_i)$ must be bounded, as $\dispersion^{\star}$ is coercive. Thus, if it does not converge to $\hat g_j(Z_i)$ or equivalently its limit
$a:=\inner{\hat L^\star, \delta_{Z_i}}$, it must have a convergent subsequence 
$g_{j_k}(Z_i)$ that converges to $b \neq a$. Because $\dispersion^{\star}$ is continuous, it follows
that the term corresponding term in Equation~\ref{eq:midpoint-convex-d} converges to 
$\dispersion_i^{\star}(a) + \dispersion_i^{\star}(b) - 2 \dispersion_i^{\star}\{(a+b)/2\}$, which is strictly positive because $\dispersion^{\star}$ is strictly convex. It follows that along this subsequence, the whole expression in Equation~\ref{eq:midpoint-convex-d} is bounded away from zero, 
which, as before, cannot be the case if $g_j$ and $\hat g_j$ are both minimizing sequences for $d$.
\end{proof}

\printbibliography[title=Appendix References]
\end{refsection}
\fi

\end{document}